\documentclass[12pt]{article}

\usepackage{amsmath}
\usepackage{mathtools}
\usepackage{booktabs}
\usepackage{threeparttable}
\usepackage{graphicx}
\usepackage{caption}
\usepackage{float}
\usepackage{url}

\usepackage[table]{xcolor}
\usepackage{colortbl}
\usepackage{booktabs}

\usepackage[hang, flushmargin]{footmisc}
\usepackage[colorlinks=true, allcolors=black]{hyperref}
\usepackage{footnotebackref}

\usepackage{subcaption} 
\usepackage{adjustbox} 

\usepackage{tablefootnote}

\usepackage{geometry}
\usepackage{setspace}
\usepackage[sc]{mathpazo}

\usepackage{tikz}
\usetikzlibrary{positioning, arrows.meta}

\usepackage{natbib}
\usepackage[disable]{todonotes}

\title{\textbf{Corporate Earnings Calls and Analyst Beliefs}}
\author{Giuseppe Matera\thanks{Giuseppe Matera is at EPFL and Swiss Finance Institute. I am very grateful to my advisors, Andreas Fuster and Semyon Malamud, for precious advice and invaluable support. I also thank Federico Baldi-Lanfranchi, Andrea Della Vecchia, Francesco Celentano, Emanuele Luzzi (discussant), Luca Pagliuca for helpful comments, and seminar and conference participants at SFI PhD Workshop 2025, EPFL-Unil PhD Workshop Fall 2025. I gratefully acknowledge support from the Swiss National Supercomputing Centre (CSCS) through project ID lp85 on the Daint-Alps system.}}
\date{\today}

\begin{document}
\onehalfspacing

\maketitle

\begin{abstract}

Economic behavior is shaped not only by quantitative information but also by the narratives through which such information is communicated and interpreted \citep{Shiller2017}. I show that narratives extracted from earnings calls significantly improve the prediction of both realized earnings and analyst expectations. To uncover the underlying mechanisms,  I introduce a novel \textit{text-morphing} methodology in which large language models generate counterfactual executive presentation from earnings calls that systematically vary topical emphasis (the \textit{prevailing narrative}) while holding quantitative content fixed. This framework allows me to precisely measure how analysts under- and over-react to specific narrative dimensions. The results reveal systematic biases: analysts over-react to sentiment (optimism) and under-react to narratives of risk and uncertainty. Overall, the analysis offers a granular perspective on the mechanisms of expectation formation through the competing narratives embedded in corporate communication.

\end{abstract}






\clearpage

\section{Introduction}

Economic decision-making is shaped not only by data and models, but also by the narratives through which agents interpret and communicate economic reality. As \cite{Shiller2017} argues, narratives---compelling stories that convey meaning and emotional resonance---spread across economic actors much like contagions, influencing behavior and expectations in ways that traditional models of rational information processing often overlook. Yet, despite their acknowledged importance, the empirical content of such narratives remains difficult to quantify, and their role in shaping belief formation is not well understood.\footnote{Studies using surveys and large language models to examine and measure the role of narratives in financial markets and the macroeconomy include \citet{Andre2022}, \cite{Bybee2023}, \cite{Barron2023}, \cite{Andre2023}, \cite{Dim2023}, \cite{Flynn2024}, \cite{Andolfatto2025}.}

This paper offers an empirical approach to this problem by focusing on a setting where narratives are explicitly produced and transmitted: the interaction between corporate managers and financial analysts. Financial analysts occupy a central position in this narrative ecosystem. Tasked with interpreting corporate disclosures and market information, they act as intermediaries between firms and investors, translating managerial communication into forward-looking assessments of firm value. While the full-information rational expectations (FIRE) framework posits that forecasts should reflect the unbiased and comprehensive use of quantitative data, substantial evidence indicates that analysts' expectations systematically deviate from this benchmark. Analysts tend to underreact to new information in the short term \citep{Livnat2006} and overreact in the long term by extrapolating recent positive signals \citep{LaPorta1996}, with biases varying by forecast horizon \citep*{Dessaint2024, deSilva2024}. These systematic biases also relate to important macroeconomic and financial phenomena, including business cycles and asset price variation.\footnote{See \cite{Bondt1985}, \cite{Vissing-Jorgensen2003}, \cite{Bacchetta2009}, \cite{Malmendier2011}, \cite{Kelly2012}, \cite{Greenwood2014}, \cite{Bordalo2019, Bordalo2024, Bordalo2025}, \cite{Balakrishnan2014}, \cite{Giglio2021}, and \cite{Nagel2022}, among others, and \cite{Adam2023} for a review.}

These empirical regularities suggest that analysts belief formation processes are influenced by more than hard data alone. Consequentially, framing, tone, and emphasis, the narrative elements of corporate communication, play a crucial role in shaping how information is interpreted and transmitted. Earnings calls provide a particularly suitable context for this research. Held quarterly, they serve as both disclosure and performance overview: occasions where executives present quantitative results while also framing the firm's trajectory through interpretive narratives. For analysts, these events offer a concentrated flow of both hard and soft information, with forecast revisions peaking in the immediate aftermath of the call. As such, earnings calls constitute a natural laboratory for testing how narratives are constructed, transmitted, and internalized within financial markets—and whether the attention financial analysts devote to them is justified by their informational content.


The first contribution of this paper is to examine whether and how analysts incorporate information expressed and framed in corporate earnings calls into their belief formation processes. I also study whether analysts' focus on these narratives reflects genuine informational value or merely responds to managerial framing. I show that the narrative and linguistic content of earnings calls indeed contains incremental information about firms' fundamentals, and analysts take this into account while forming their beliefs. Extracting such narratives requires modern Large Language Models (LLMs) because corporate disclosures are inherently high-dimensional and context-dependent, and early textual proxies, such as word counts or readability indices, fail to capture the semantic nuance and communicative style. 



Specifically, to represent text, I compute earnings call transcripts embeddings\footnote{Embeddings are high-dimensional vector representations of the text capturing semantic and syntactic features. Every LLM generates a different embedding of a given text.} using a finance-specific FinBERT model pretrained on financial text available before 2014.\footnote{Implementation details appear in later sections. For the model, see \url{https://huggingface.co/ProsusAI/finbert}.} I deliberately avoid the latest foundation models to eliminate the risk of a look-ahead bias.\footnote{Several papers (see, e.g., \cite{Glasserman2023}, \cite{Sarkar2024}, \cite{Ludwig2025}, have criticized using modern LLMs for economic research due to their inadvertent exposure during pretraining to economic data or contemporaneous news in the sample that the models are used to explain. By contrast, all results in this paper are free from such exposure. Hence, they should be interpreted as conservative: more expressive, state-of-the-art models would likely capture additional meaning and structure in the transcripts, yielding potentially stronger findings than those reported in this work.} 

To pin down the incremental information contained in the narratives, I benchmark my results against a comprehensive set of more than 300 numerical firm characteristics constructed from the Global Factor Data of \citet{Jensen2023} (henceforth referred to as JKP data), as well as the most recent financial statements from the SEC 10-Ks filings. I also include abnormal stock returns (FF5-adjusted) on the earnings announcement date, along with realized earnings yields and the Standardized Unexpected Earnings (SUE; \citep{Livnat2006}), as additional characteristics. 


I estimate several machine learning (ML) models based on three alternative feature sets: one using numerical stock characteristics alone, one using embeddings of earnings call transcripts alone, and one combining both sources. I then evaluate the models' out-of-sample performance primarily through changes in the out-of-sample R-squared. Since direct comparisons of R-squared values can be sensitive to differences in the variability of test samples, I complement this analysis with the \cite{Clark2007} test to assess whether the observed improvement in predictive accuracy is statistically significant. Textual information derived from earnings calls markedly improves the prediction of analysts' expectations and forecast dispersion. Incorporating these linguistic measures provides a statistically significant enhancement in capturing analysts' beliefs, beyond what can be explained by traditional quantitative predictors. These results suggest that linguistic content from managerial disclosures provides incremental information beyond traditional quantitative fundamentals, capturing nuanced aspects of firm communication that complement established information sources in the analysts' forecast formation process.

The second contribution of this paper is to identify which narratives of corporate communication are most influential in shaping analysts' beliefs, and whether these same features predict subsequent realized earnings.

Estimating what causes particular analysts' reactions would, in an ideal world, require rerunning the same earnings call while varying only one clearly identified narrative of the CEO's remarks---an experiment which is clearly infeasible. Instead, I propose a novel framework that (i) constructs credible counterfactual disclosures: precise, transparent edits (henceforth, morphs) to the CEO's speech along pre-specified linguistic dimensions; (ii) holds fundamentals and the market's information set fixed; and (iii) measures how each \textit{morph} changes analysts' responses. 


I identify six key narrative dimensions likely to influence analysts' belief formation: Guidance, Jargon, Confidence, Global Perspective, Sentiment, and Uncertainty. These dimensions are selected based on prior research on the effects of linguistic features and stock market reactions. To verify that these narratives are informative, I use a state-of-the-art large language model to assign each earnings call a score from 0 to 10 for the extent to which each narrative dimension is expressed. A ML model trained on these six scores achieves an average out-of-sample performance of approximately $40\%$ relative to models using the full textual embeddings, suggesting that the six narratives capture informational content relevant to analysts' belief formation.

Then, inspired by the methodology of \cite{Horton2023} and \cite{Ludwig2024}, I construct an \textit{in-silico} experimental design with a simple approach to generate human-like text. Namely, I train a model that maps earnings call transcripts to analysts' measurable actions; then I create systematic, structured, human-like edits (morphs) to the same transcript along the above-mentioned six linguistic dimensions. Finally, I generate embeddings of the morphed text and feed them into my ML model. The difference between the outcomes based on the original (baseline) and morphed (counterfactual) texts recovers the predicted treatment effect (PTE) of the targeted language narrative. This methodology advances beyond conventional natural-language-processing framework that relies on simplistic text-based proxies (e.g., bag-of-words, fixed sentiment lexicons) by making it possible to intervene directly on the text. The granularity and flexibility of this framework make it possible to quantitatively evaluate the influence of narratives within financial markets and to estimate with precision the contribution of each narrative dimension to realized earnings and to analysts' earnings expectations.

Most importantly, this \textit{in-silico} experiment allows me to disentangle the narratives that financial analysts pay attention to---those driving the ML model that predicts their forecasts (Analysts Expectation for Earnings, AEE)---from the narratives they \textit{should} pay attention to---those associated with future realized earnings (Future Realized Earnings, FRE). I find that, while analysts appear to allocate a rational level of attention to jargon, they severely under-react to narratives of uncertainty (Average PTE for AEE = -9.22 bps versus Average PTE for FRE = -41.09 bps), confidence (Average PTE for AEE = 6.09 bps, Average PTE for FRE = 25.76 bps), and future guidance (Average PTE for AEE = 10.84 bps, Average PTE for FRE = 19.43 bps). Conversely, they over-react to narratives of sentiment (Average PTE for AEE = 34.88 bps, Average PTE for FRE = 25.76 bps) and global perspective (Average PTE for AEE = 10.84 bps, Average PTE for FRE = -19.43 bps).



To summarize, I show that language, although it conveys no additional quantitative information and often reflects managerial framing rather than substantive disclosure, is still informative about future economic prospects and can induce sizeable shifts in analysts' forecasts. This finding underscores the powerful role of corporate communication in shaping market expectations and, more broadly, the role of narrative within financial markets. The framework developed in this paper provides a way to precisely measure the effects of such narratives and sheds new light on the mechanism of expectation formation within financial markets. For corporate managers, it offers an opportunity to strategically design communication by experimenting with alternative framings of the same message. For analysts, it highlights the importance of distinguishing substantive information from rhetorical choices and of redirecting attention away from topics that may be emphasized for strategic rather than informational purposes.

\subsection*{Related Literature}

This paper connects the literature on subjective belief formation with the emerging research that leverages large language models (LLMs) for social science inference. By using earnings calls as the textual input, the analysis develops experimental environments that cannot be implemented through traditional experimental designs.

\paragraph{Subjective beliefs and analyst expectations.}  
A substantial body of research shows that equity analysts form expectations under cognitive constraints and frictions, and examines the determinants and consequences of these limitations. This literature includes studies that analyze quantitative forecasts obtained from surveys or elicited through the traditional Institutional Brokers Estimates System (I/B/E/S). (\cite{Easterwood1999}, \cite{Diether2002}, \cite{Livnat2006}, \cite{Greenwood2014}, \cite{Bouchaud2019}, \cite{Ma2024}, \cite{Bordalo2019,Bordalo2024, Bordalo2025}, \cite{Giglio2021}, \cite{Bastianello2023, Bastianello2024, Bastianello2025}, \cite{Nagel2022}, \cite{Ben-David2024}, \cite{Dahlquist2024}, \cite{HDcaire2023, HDcaire2024}, \cite{DeLaO2021, DeLaO2024}, \cite{Delao2025}).

A rapidly expanding literature combining machine learning, textual analysis, and finance to study belief formation (\cite{VanBinsbergen2023}, \cite{Charles2025, Kendall2025}, \cite{Bybee2025}, \cite{Giglio2022}, \cite{Bastianello2022}, \cite{Gormsen2024, Gormsen2025}, \cite{Bianchi2024}, \cite{Gabaix2025}, \cite{Chen2024}, \cite{Chen2025}, \cite{Lopez-Lira2023}, \cite{Lv2024}, \citet{Cohen2024}, \cite{Sarkar2025}, \cite{Stolborg2025}).\footnote{See \cite{Kothari2016} and \cite{Giglio2025} for a more extensive review.}

Recent work has increasingly used sell-side equity analyst reports as a primary data source. For example, \cite{DeRosa2024}, \cite{Ke2024}, and \cite{Bastianello2024} combine these reports with I/B/E/S forecast data to study different dimensions of analysts' information processing, including memory, attention, and the structure of their mental models.

In this paper, I take a complementary approach and examine analysts' expectation-formation processes using corporate earnings calls, which allows me to measure how the information environment of the corporate earnings calls shapes analysts' beliefs.

\paragraph{LLMs as tools for empirical social science.}
Recent work shows that large language models can reliably reproduce classic behavioral patterns, simulate experimental decision-making environments, and act as transparent proxies for human judgment (\citet{Horton2023}, \cite{Hansen2024}, \citet{Manning2024}, \cite{Bhagwat2025}, \citet{Ludwig2025}, \cite{Kazinnik2025}). They can also implement causal inference through textual interventions, emulate survey-based expectation-formation exercises, and capture heterogeneity in how different \textit{in-silico} ``agents'' interpret the same information. Together, these findings demonstrate the potential of LLMs to complement or substitute traditional experimental and survey methods in social science research.

\paragraph{Financial disclosures and textual analysis.}
This paper also builds on research examining the informational role of corporate narratives. A growing body of work analyzes how different forms of financial disclosures shape market participants' beliefs and decisions. Studies using corporate earnings calls, including \cite{Dzieliski2017}, \cite{Hassan2019}, \cite{Cohen2020}, \cite{Li2021}, \cite{Mamaysky2023}, \cite{Meursault2023}, \cite{Sautner2023}, \cite{Cohen2024}, and \cite{Siano2025}, investigate the linguistic and tonal features of managers' communication and their impact on investor reactions. Complementary research on 10-K filings, such as \cite{Loughran2011}, \cite{Hoberg2010}, \cite{Hoberg2015}, \cite{Hoberg2016}, and \cite{Kim2024}, explores how narrative content and disclosure structure contribute to the transmission of firm-specific information. Together, these studies highlight the central role of corporate communication in shaping how information is interpreted in financial markets.\footnote{See \cite{Hoberg2025} for a more extensive review of papers and future directions.}

\paragraph{Contribution.}  
By integrating these strands, this paper proposes a framework in which LLMs are used to construct counterfactual versions of earnings calls, designed to test how specific linguistic changes affect analysts' belief updating. In doing so, the paper contributes to both the behavioral finance literature on subjective expectations and the methodological literature on LLMs as instruments for social-scientific experimentation.

\paragraph{Outline.} 
Section~\ref{sec:data} describes the data, including the construction of text-based features and the 
definition of the prediction tasks. Section~\ref{sec:method} illustrates the novel methodology. Sections~\ref{sec:prediction_res} and~\ref{sec:counterfactual_res} present the empirical results on the 
prediction of analysts' beliefs and on the treatment effects of counterfactual earnings 
calls, respectively. Section~\ref{sec:conclus} concludes.

\section{Data}
\label{sec:data}

Annual earnings announcements are scheduled corporate events through which firms disclose fundamental information to the market, most notably the earnings figures that give the event its name. On such occasions, publicly traded firms typically hold earnings calls, during which senior management reports recent financial results and provides forward-looking guidance to investors and financial analysts. It has also become increasingly common for firms to file their annual 10-K reports with the SEC within two days of the earnings release \citep{Arif2019}. 

Trying to predict the future, financial analysts incorporate a large number of data and information sources, use complex models and engage in corporate interactions. In this paper, I assume that the financial statements filed in these 10-Ks enter analysts' information set and shape their expectations about the future performance of the company. In addition, I control for standard stock-level characteristics, obtained from the \cite{Jensen2023} dataset (henceforth the JKP dataset), to measure the predictive contribution of earnings call content, beyond information that may be contained in accounting and market-based firm characteristics.

\subsection{Corporate Earnings Calls}

I obtain earnings conference call transcripts from Capital IQ–Transcripts, which provides verbatim records of corporate and institutional events. I retain all English-language transcripts related to US public companies that announce their annual earnings. Although I use the full history of available calls, coverage before 2008 is sparse and relies mostly on other sources (e.g., SeekingAlpha). My sample comprises 35,203 earnings call transcripts corresponding to annual earnings announcement days, for a total of 5,370 companies, covering the full set of forecast horizons considered in this study. Details on the dataset are in Section~\ref{section:latest_version_transcripts} in the appendix.

Earnings calls typically consist of two distinct sections: a prepared statement by executives, followed by a question-and-answer (Q\&A) session that allows investors and financial analysts to comment, pose questions, and potentially challenge management's remarks. The Capital IQ data are available at the speaker level, allowing the two segments of each earnings call—i.e., the prepared remarks from management and the Q\&A session—to be distinguished. For the rest of the analysis, I focus only on the management remarks and how these affect analyst expectation formation.

\begin{figure}[htbp]
\centering
\caption{\textbf{Earnings Calls: Number per Year and Average Length}}
\caption*{This figure illustrates the combined dataset of corporate earnings calls and associated analyst forecasts. The blue bars (left axis) represent the total number of earnings calls held each calendar year, while the red line (right axis) indicates the average number of words in the management remarks section per call. The sample includes only earnings calls followed by at least one individual analyst forecast, across all forecast horizons. Each earnings call typically consists of two parts: the prepared management remarks and the question\&answer session with analysts and investors. Only earnings call transcripts containing non-empty management remarks are retained in the dataset.}
\label{fig:corporate_calls_hist}
\includegraphics[width=0.9\textwidth]{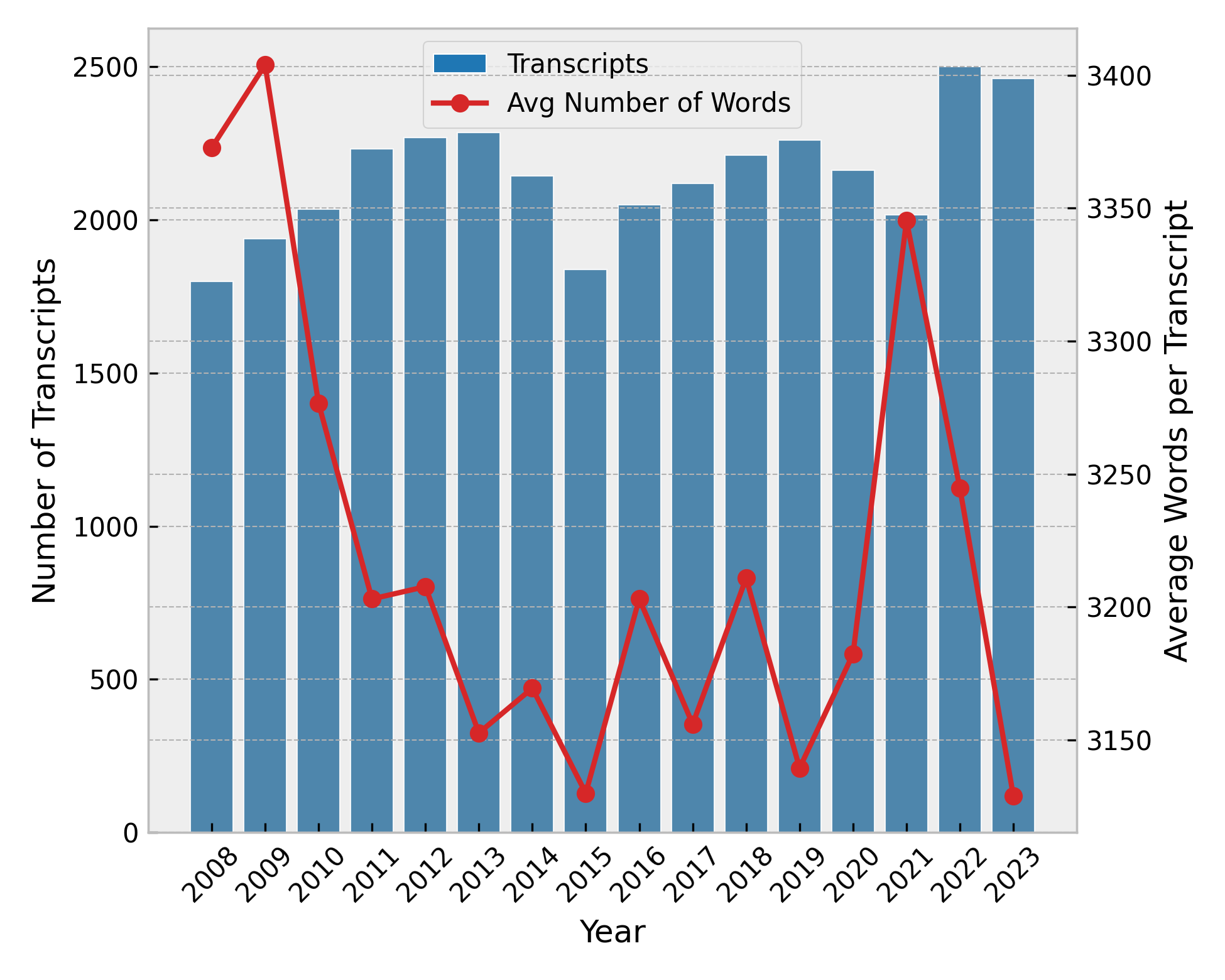}
\end{figure}

Figure \ref{fig:corporate_calls_hist} shows that, over the sample period, the number of earnings calls related to annual earnings announcements is stable at roughly 2,000 per year. The structure of these calls is likewise stable, with an average length of about 3,000 words for the management speeches. The only two notable deviations occur around the 2008 Financial Crisis and the outbreak of the Covid-19 crisis, when these remarks were, on average, about 10\% longer.

\begin{figure} 
\centering
\caption{\textbf{Timing of Analyst Forecasts Around Earnings Calls}}
\caption*{
This figure depicts the distribution of analyst forecast releases in the weeks surrounding earnings calls. It reports the weekly proportion of individual analyst forecasts with annual periodicity, across all forecast horizons, relative to the total number of forecasts issued within the seven-week window around the call. Week 1 corresponds to the week beginning on the day of the earnings call. Since earnings calls occur quarterly, forecasts issued more than seven weeks before or after a call are associated with the preceding or subsequent event. The data cover the period from 2008 to 2023.
}
\label{fig:event_study}
\includegraphics[width=0.8\textwidth]{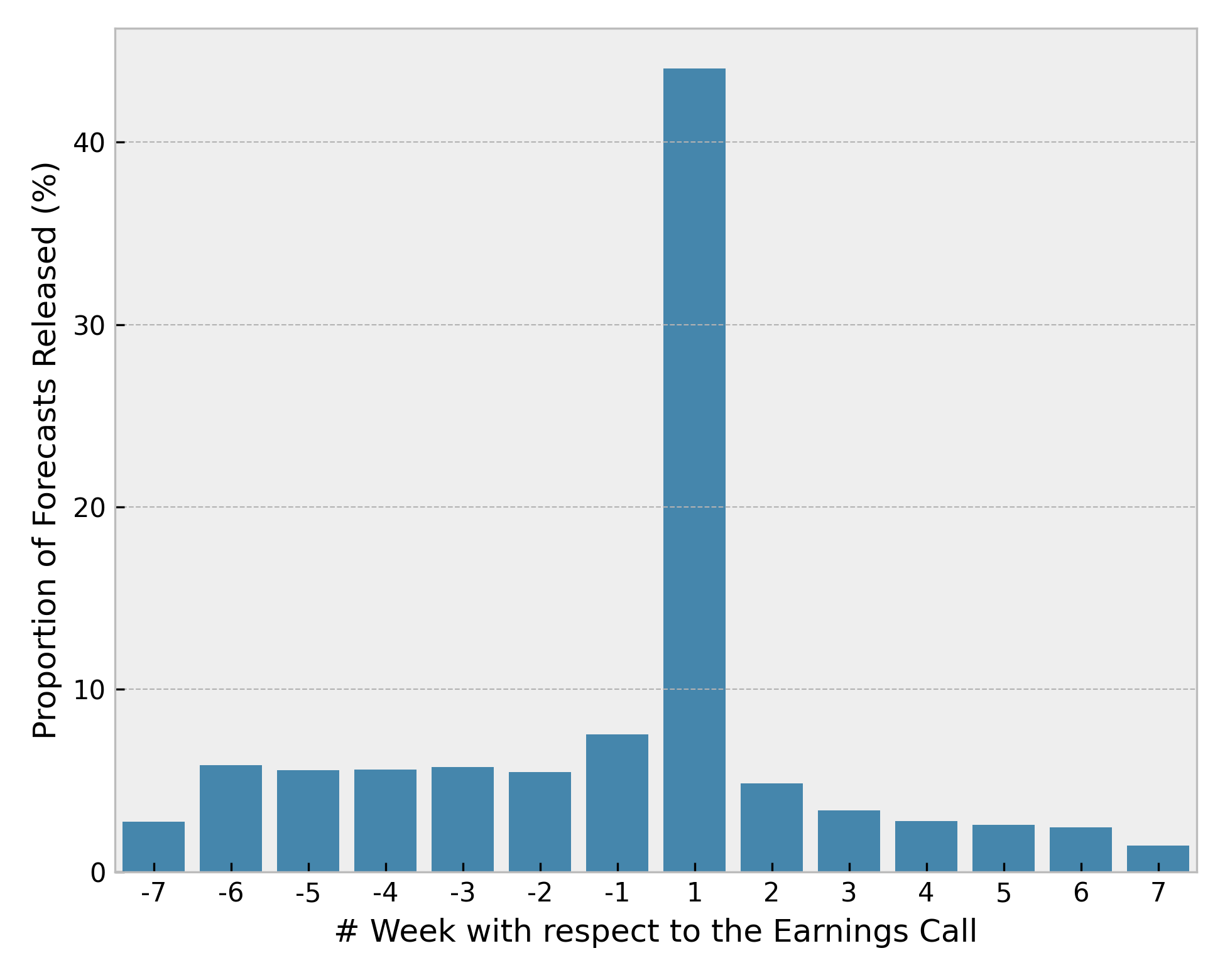}
\end{figure}

Immediately following these days, the volume of financial analyst reports surges, reflecting the assimilation and processing of the newly released information, as illustrated in Figure \ref{fig:event_study}.

\subsection{Analyst Forecasts}

I construct the sample of analyst expectations using analyst-level earnings-per-share (hereafter EPS) forecast data from the Institutional Brokers' Estimate System (IBES) Detail Unadjusted History database, which contains earnings forecasts submitted by sell-side analysts. For the purposes of my analysis, I restrict the sample to U.S. companies and focus on one-, two-, and three-year-ahead forecasts. The IBES Detail Unadjusted History dataset reports analyst–firm–date-specific forecasts and realizations without retroactive adjustments for corporate actions such as stock splits.\footnote{Evidence suggests that concerns about other retroactive changes and ID reshuffling are relatively minor in U.S. data, indicating that the dataset can serve as a reliable proxy for market expectations. See \cite{Law2023} for a comprehensive discussion of these limitations and their empirical relevance.} These corporate actions can introduce mechanical discrepancies between forecasts and realizations \citep{Diether2002, Bouchaud2019}. To address this, I adjust all EPS forecasts and realized EPS values to the prevailing share count to ensure comparability. I obtain cumulative share adjustment factors (\texttt{CFACSHR}) from the CRSP daily stock file (\texttt{dsf}), which I merge with IBES using the WRDS IBES–CRSP linking table (\texttt{ibcrsphist}). Following this merge, I retain only observations with a valid link at the time of the forecast announcement. When the forecast date falls outside the CRSP trading calendar, I use the adjustment factor from the closest prior trading date. Finally, I impose additional filters to ensure data quality, such as correcting for temporal inconsistencies between forecast announcement dates and other relevant dates.\footnote{See \url{https://wrds-www.wharton.upenn.edu/pages/support/support-articles/ibes/anndats-actdats-some-situations-where-not-case/}
for examples.}

Following standard practice in the literature \citep{Bouchaud2019}, I use earnings yields, i.e., earnings-to-price ratios, rather than raw EPS values. To avoid look-ahead bias, I use the price prevailing at the time of the earnings call. This scaling accounts for heterogeneity in firm size and price levels, making comparisons across firms and over time more meaningful. Nevertheless, for simplicity, I continue to refer to all variables of interest as Earnings.

Summary information on the number of firms, of brokers and analysts in the final sample per horizon are reported in Table \ref{tab:fpi_counts}. As expected, broker coverage is incomplete across forecast horizons: most brokers issue near-term forecasts, while only fewer provide projections at longer horizons. Consequently, observations thin out as the horizon extends. All industries are included in the baseline sample, as the general perspective used in this paper adopts measures which are normalized using each firm's past accounting figures as a benchmark. 


\subsection{Matching Forecasts to Earnings Calls}

I link analyst forecasts to conference calls by matching Capital IQ transcripts with forecast data from the IBES Unadjusted Detail file, using the linking tables provided by WRDS. 

To capture the effect of the flow of information present in the earnings call on analysts' expectation-formation process, I focus on all forecasts issued within 15 days of the call date, thereby reducing the risk that subsequent news contaminates their forecasts. I construct the consensus forecast by computing the median of all available forecasts, rather than relying on the consensus provided by the data vendor, since the latter may incorporate forecasts outside the event window.

I exclude earnings call transcripts that are associated with multiple companies, as such cases could create ambiguous links between the text and firm-specific outcomes and potentially induce spurious predictability.

\subsection{Accounting and Market-Based Information}

As a starting point, I incorporate accounting and market-based information, which reflect both realized fundamentals and market perceptions, in order to measure the incremental predictive power of textual features\footnote{In the machine learning literature, the outcome and predictors are typically referred to as the "target" and "features." I adopt this nomenclature throughout this paper.}.

\subsubsection{Jensen-Kelly-Pedersen (JKP) stock characteristics}
The first purpose of this work is to compute the predictive value of earnings call transcripts beyond any hard information contained in financial statements or available to financial analysts from the stock market. While modeling such a benchmark is impossible—I cannot know what analysts actually know or think—I assume their information set is largely captured by the set of market- and accounting-based stock characteristics from the JKP dataset \citep{Jensen2023}. This dataset includes 153 characteristics. Following the taxonomy in \citet{Jensen2023}, I group firm characteristics into 13 clusters: Accruals, Debt Issuance, Investment, Leverage, Low Risk, Momentum, Profit Growth, Profitability, Quality, Seasonality, Short-Term Reversal, Size, and Value. These clusters provide a structured representation of the large cross-section of characteristics used in asset pricing studies.

I merge this dataset so that all conditioning variables reflect information available in the month preceding the conference call. In addition to the levels of these characteristics, I compute one-year changes. 

\paragraph{Standardized Unexpected Earnings (SUE).}
The Standardized Unexpected Earnings (SUE) is a measure of earnings surprise, defined as the deviation of reported earnings per share from the prevailing analyst consensus prior to the release. The JKP data already include SUE, but because I merge the dataset using information from the month before the earnings call, the variable refers to the prior quarter rather than the quarter of interest.Therefore, I recompute SUE from the IBES data. Following \cite{Livnat2006}, I use only estimates issued within 90 days before the earnings call date, retaining the latest forecast when an analyst issues more than one in order to avoid stale information. I then normalize the resulting earnings surprise by the stock price at the report date.

\subsubsection{Financial Statements}

Financial Statements (FS) are a central input in analysts' work, as they reveal trends in accounting information and provide the basis for constructing financial ratios. In my analysis, I include the three main statements filed by firms in their 10-Ks: the Income Statement, Balance Sheet, and Statement of Cash Flows.\footnote{In order to retrieve Balance Sheet, Income Statement and Statement of Cash Flows, I use the query available at \url{https://wrds-www.wharton.upenn.edu/pages/get-data/compustat-capital-iq-standard-poors/compustat/north-america-daily/simplified-financial-statement-extract/}.} I match these with the dataset described above using company identifiers and fiscal year-end dates. In addition to the levels of these variables, I compute one-year changes, as I do for the JKP characteristics.

Because these variables differ in scale, distribution, and time variation, I rank-standardize them within each date to the \( [-0.5, 0.5] \) interval before merging. This procedure places all variables on a common, bounded scale that is robust to outliers and time-varying dispersion. 


\section{Methodology}
\label{sec:method}

\subsection{Analyst Expectation Formation around Earnings Calls}

Analysts are more active in issuing forecasts immediately after earnings calls. Figure \ref{fig:event_study} shows a sharp increase in the issuance of analyst forecasts immediately after the earnings call. Prior studies suggest that this surge reflects analysts processing newly released financial statements and earnings information. In this paper, I argue that the way management frames information, and the qualitative guidance provided during earnings calls, matter over and above the quantitative disclosures. 

To do so, I study how earnings-call content affects analyst forecasts, restricting attention to forecasts issued within 15 calendar days after the earnings call to minimize interference from subsequent news. I use a 15-calendar-day period after the earnings call to capture the the majority of forecasts related to that precise event, while limiting exposure to unrelated news. I then construct a consensus forecast by taking the median of all forecasts within this window. 

In the analysis, following \cite{Bouchaud2019}, I use the earnings yield (earnings per share divided by stock price), which facilitates comparisons across firms relative to raw EPS. To avoid look-ahead bias, I use the stock price prevailing at the time of the earnings call. Nevertheless, for simplicity, I continue to refer to the forecasted and realized variable as earnings.

I track how analysts adjust their beliefs along three dimensions following an earnings call. First, I examine the median expected change in analysts' forecasts, a robust aggregate measure that summarizes how expectations shift. Second, I assess whether this consensus tends to overshoot or undershoot the earnings that are ultimately realized, signaling the presence of under- or over-reaction. Third, I consider the cross-sectional standard deviation of forecasts in the consensus to gauge the degree of disagreement among analysts. Together, these measures describe not only the central movement in beliefs but also their accuracy and the degree of consensus.

For clarity, Figure \ref{fig:timeline} summarizes the earnings-release timeline. At date \(t\), the firm reports results for the prior period, \(Y_{t-1}\), and holds the earnings call. Analysts then issue (or revise) forecasts for \(Y_{t+h}\) within the subsequent 15-calendar-day window. At date \(t+h\), the firm reports realized earnings \(Y_{t+h}\).

To describe the variables related to analysts behavior, let me assume at $t$, a company $i$ releases 
\[
y_{i,t} \coloneqq \text{ the earnings for the previous period}, Y_{t-1}, \text{ released at } t.
\]
and holds an earnings call with financial analysts and investors. Then, the analysts issue forecasts for a set of forecast horizon, incorporating information just heard. Their forecasts are aggregated forming a consensus estimate,
\[
F_{i,t}(y_{i,{t+h}}) \coloneqq \text{ consensus formed at } t \text{, for the earnings released at } t+h.
\] 

During the earnings call, the newly realized earnings becomes public and enters analysts' information sets. As a naive benchmark---abstracting from any private information---there is no reason to expect the future earnings to deviate from the most recently realized level. To characterize departures from this benchmark, Eq.~\ref{eq:expected-change} defines the \textbf{Analysts' Expected Change in Earnings}, capturing how analysts believe the future value will differ from the current one. For each firm $i$ and horizon $h$, I define

\begin{equation}
\label{eq:expected-change}
\underbrace{EC_{i,t,h}}_{\text{Expected Change in Earnings}}
= F_{i,t}(y_{i,{t+h}}) - y_{i,t}.
\end{equation}

Individual financial analysts naturally form their own expectations about future earnings, even when exposed to the same information during a corporate earnings call. Such dispersion can arise from heterogeneous priors, risk tolerances, models, and, crucially, differences in how the same language is interpreted. Therefore, I examine cross-analyst disagreement using the dispersion of expected changes, defined in Equation \ref{eq:dispersion-change} for a firm $i$ and a forecast horizon $h$, and test whether specific linguistic features are systematically associated with higher or lower disagreement in expectations. To ensure comparability across firms and consistency with other variables here defined, I scale the cross-sectional dispersion of forecasts by the stock price prevailing at the earning call date and define \textbf{Forecast Disagreement} as 

\begin{equation}
\label{eq:dispersion-change}
\underbrace{\frac{\sigma_{EC_{i,t,h}}}{P_{i,t}}}_{\text{{Forecast Disagreement}}}
= \operatorname{St. Dev.}\!\big(EC_{i,t,h}\big)
\end{equation}



Although corporate earnings calls do not mechanically determine these realizations, I estimate a model predicting the realized change in earnings to identify which narratives in corporate communications are most closely related to realized fundamentals. This, in turn, allows me to construct an \textit{in-silico} counterfactual through which I can study how analysts form their beliefs. I define \textbf{Realized Change in Earnings} as follows:

\begin{equation}
\label{eq:realized-change}
\underbrace{\Delta E_{i,t,h}}_{\text{Realized Change in Earnings}}
= y_{i,t+h} - y_{i,t}
\end{equation}

All the target variables are trimmed at the 5\% and 95\% levels. The summary statistics for the afore-defined variables can be found in Table \ref{tab:summ-stats}.

\subsection{Representation in Textual Spaces: Earnings Call Embeddings}

Text is inherently high-dimensional and context-dependent. To obtain compact numeric representations that preserve semantics and tone, I use textual embeddings.

Textual embeddings are continuous, fixed-length vector representations of text that place words, sentences, or documents in a geometric space where semantic similarity corresponds to proximity. The geometry of this space captures nuances such as synonymy, tone, and topical relatedness, providing a compact and meaningful summary of the original text.

State-of-the-art model embeddings are produced by transformer encoders trained on large corpora with self-supervised objectives; they yield contextual representations in which the same word can map to different vectors depending on its surrounding context. 

Creating a framework able to predict future analyst forecasts from earnings call transcripts presents two serious concerns: text length feasibility and look-ahead bias
.
Earnings calls are long and often exceed transformer input limits, which can make downstream tasks (classification, similarity) sensitive to how long texts are chunked and aggregated. To mitigate this, I focus on the management's prepared remarks, the most structured and information,  dense segment of each earnings call.

To represent each earnings call as a single vector, I compute textual embeddings on non-overlapping segments of the transcript that fit the encoder's context window. The resulting segment-level vectors must be combined to form a document-level representation. There is no consensus on a uniquely superior pooling rule for long texts, and--consistently with prior practice--reasonable alternatives perform similarly in this setting. I therefore employ an approach that preserves as much information as possible, using the widely used Hugging Face \texttt{transformers} library: within each segment, I average the token representations across all the tokens, from the last transformer layer, and then aggregate the segment vectors to obtain embeddings for each earnings call transcript. This approach retains signal that is distributed across all the earnings call; results are qualitatevily similar under common alternatives such as mean, max, or \texttt{[CLS]}-based pooling.

Text models can ``peek into the future'' if \emph{content leakage} lets them read explicit future cues (dates, magnitudes), or \emph{temporal leakage} lets them learn from documents that occur after the prediction target. Either failure mode would inflate performance and weaken the credibility of my counterfactual exercise. \citep{Sarkar2024}. 

\paragraph{Masking.} 
A threat to the downstream tasks in this article would involve managers potentially mentioning metrics and accounting figures during presentation in earnings calls, making the soft information and the framing of the latter unidentifiable with textual embeddings.

To address these issues, I mask \emph{all} numerals e.g., accounting figures, guidance ranges, percentages, dates, quarters/years, and ticker-like digit strings--masking them using the model's \texttt{[MASK]} token provided by the Hugging Face tokenizer.\footnote{For reference, see \url{https://huggingface.co/docs/transformers/en/main_classes/tokenizer}.} This forces the models to rely on semantics, tone, and discourse structure rather than magnitudes or calendar markers (e.g., terms like ``\emph{Q4 2022}'' or ``{+}35\%''), and it prevents subtle forward references (such as ``updated 2027 outlook'') from acting as proxies for realized outcomes.

I also use a finance-specific pretrained encoder model i.e., \href{https://huggingface.co/ProsusAI/finbert}{ProsusAI/FinBERT}, whose training data extends to 2017, which I do not fine-tune on my earnings calls dataset. Furthermore, this model was trained on general-domain and financial news corpora, not on earnings-call transcripts, and its upstream training data predate my evaluation windows.

This design attenuates the risk of look-ahead bias, where future information would otherwise hamper prediction.

\section{Earnings Calls and Analyst Forecasts}
\label{sec:prediction_res}

Analysts are continuously exposed to a wide range of information. In forming their forecasts, they routinely incorporate market-based information, balance sheets and other financial statements into their models. A central question, however, is whether the corporate communication conveyed during earnings calls also influences analyst expectations. If it does, does this communication merely complement existing numerical data---acting as a proxy for information already available---or does it provide independent and additional predictive content?

\subsection{Predicting Analyst Beliefs with Machine Learning}

To address these questions, I train and evaluate two model specifications. The first (i) is a stock characteristics-only model, which uses the JKP dataset together with the most recent financial statements (in levels and one-year changes), the prevailing abnormal stock return at the earnings call date, and the most recent Standardized Unexpected Earnings. The second (ii) augments this baseline with textual embeddings extracted from earnings call transcripts. The two specifications comprise 446 and 1,214 features, respectively.

I employ a Gradient Boosting regressor, a machine-learning model which accommodates non-linearities and interactions while offering competitive computational performance. The analysis is conducted separately by forecast horizon, in order to assess whether the contribution of transcript-based textual features is concentrated in the near term or persists at longer horizons.

For each forecast horizon and target variable, the most recent 30\% of observations are held out as an out-of-sample test set, while the remaining 70\% are used for model fitting and hyper-parameter tuning. To avoid overfitting and to ensure robustness of the selected hyper-parameters, I implement $K$-fold cross-validation within the training set. 

Because all numerical values and dates are comprehensively masked, the text features contain no forward-looking cues, eliminating content-based look-ahead. In addition, the test data for these predictive exercises are well after the last time stamp of the FinBERT training sample, ensuring that the embeddings are applied strictly out-of-sample.

Out-of-sample performance is evaluated using two standard metrics: the Mean Squared Error (MSE) and the R-squared statistic. The MSE measures the average squared difference between predicted and actual values, as defined in Eq.~\ref{eq:mse}, while the R-squared captures the share of variance in the test set explained by the model relative to the test-set mean, as reported in Eq.~\ref{eq:r2}.  

\begin{equation}
\label{eq:mse}
\text{MSE}(y, \hat{y}) = \frac{1}{n} \sum_{i=1}^{n} (y_i - \hat{y}_i)^2
\end{equation}

\begin{equation}
\label{eq:r2}
\text{R-squared}(y, \hat{y}) = 1 - \frac{\sum_{i=1}^{n} (y_i - \hat{y}_i)^2}{\sum_{i=1}^{n} (y_i - \bar{y})^2}
\end{equation}

Here, $\hat{y}_i$ denotes the model's prediction of the outcome $y_i$ for observation $i$, and $\bar{y}$ represents the sample mean of the dependent variable in the test set. A positive R-squared indicates that the model explains more variation in the dependent variable than a naïve benchmark that predicts only the sample mean.  

This measure is used to evaluate out-of-sample performance; for readability, I omit the ``out-of-sample'' (``OOS'') label in what follows.

\subsection{Do Financial Analysts Pay Attention to Managerial Communication?}

To evaluate whether financial analysts pay attention to managerial communication, I test whether textual features extracted from earnings calls provide incremental predictive power beyond numerical information when forecasting analysts' subsequent behavior. The idea is that, if analysts incorporate qualitative cues from management discourse into their decision-making, the textual component should convey additional information not already captured by quantitative variables.

In order to assess the incremental predictive value of textual features, I set up a simple comparative framework. 

Let $\{(y_i, x_i^{S}, x_i^{T})\}_{i=1}^{n}$ be the time-ordered sample, where $y_i$ is one of the afore-mentioned four variables capturing analysts' actions, $x_i^{S}\in\mathbb{R}^{p_S}$ are purely numeric characteristics, and $x_i^{T}\in\mathbb{R}^{p_T}$ are textual embeddings. 

I define $x_i \coloneqq (x_i^{S}, x_i^{T})$ and $X \coloneqq X_{\text{stock characteristics}} \cup X_{\text{textual embeddings}}$, and adopt a single generic modeling framework,
\[
y_i \;=\; f^{(k)}\!\big(z_i^{(k)}\big) + \varepsilon_i,
\]
and instantiate it by varying only the predictor set:
\begin{enumerate}
\item[(S)] $z_i^{(S)} = x_i^{S}$, hence $y_i = f^{(S)}(x_i^{S}) + \varepsilon_i$.
\item[(T)] $z_i^{(T)} = x_i^{T}$, hence $y_i = f^{(T)}(x_i^{T}) + \varepsilon_i$.
\item[(ST)] $z_i^{(ST)} = (x_i^{S}, x_i^{T})$, hence $y_i = f^{(ST)}(x_i^{S}, x_i^{T}) + \varepsilon_i$.
\end{enumerate}



Figure \ref{fig:just text} presents the R-squared values for models using textual features, the (T) specification , alone across different forecast horizons and outcomes namely, \emph{Expected Change}, \emph{Forecast Disagreement}, and \emph{Realized Change}. The explanatory power of text is consistently positive for expected changes and disagreement, ranging from roughly 25\% to 40\% depending on the horizon. Although the contribution slightly declines with longer horizons, the model still retains substantial predictive strength even at three years ahead.

Overall, the figure highlights the remarkable informativeness of text-based features: even without incorporating any numerical or firm-level fundamentals, the model captures a sizeable share of the variation in analyst expectations and realized outcomes. This performance underscores how linguistic signals embedded in earnings calls or reports contain rich forward-looking information that analysts and markets process over time.

This suggests that language in earnings calls not only informs near-term analyst reactions but also embeds systematic cues that relate to longer-term forecast accuracy.

Table~\ref{tab:r2_levels} summarizes out-of-sample performance across horizons. These results extend previous research by highlighting the role of highly non-linear machine learning methods and the value of incorporating an increasingly rich set of predictors. Beyond traditional accounting variables, market-based measures and analyst-related information contribute meaningfully, with unstructured inputs like textual embeddings yielding incremental predictive performance.
\citep{Chen2022}\footnote{Quoting from \cite{Chen2022}: ``We use two machine-learning methods to predict the level of earnings and the amount of earnings changes following the same rolling windows as in Subsection~3.2.3. Consistent with prior research, we observe a low out-of-sample R-squared of 5.3\% (6\%) for random forests (stochastic gradient boosting) in predicting the level of one-year-ahead earnings, lower than the out-of-sample R-squared of 7.5\% for a simple random-walk model. We also observe a low out-of-sample R-squared of 8\% (5.8\%) for random forests (stochastic gradient boosting) in predicting the amount of one-year-ahead earnings changes.''}

\paragraph{Empirical Test of Analyst Attention to Managerial Communication.}

To further validate whether the inclusion of textual features enhances predictive performance relative to the characteristics-only benchmark, I apply the MSPE-adjusted test for nested models developed by \citet{Clark2007}. This approach is particularly suitable in settings where models are nested and traditional measures such as out-of-sample R-squared may be noisy or difficult to interpret in small samples. 

The test compares the squared forecast errors of the competing models but adjusts for the mechanical disadvantage faced by the larger, nested specification: additional estimated parameters can inflate prediction error even when the extra information is genuinely useful. The adjustment removes this noise-induced penalty and evaluates whether the average difference in accuracy is positive, with the resulting test statistic defined in Eq.~\ref{eq:cw}.

\begin{equation}
\label{eq:cw}
CW = \frac{\bar{\tilde{d}}}{\sqrt{\widehat{\mathrm{Var}}(\tilde{d}_{t+1})/T}},
\qquad
\bar{\tilde{d}} = \frac{1}{T}\sum_{t=1}^T \left[ (e_{t+1}^{(r)})^2 - (e_{t+1}^{(u)})^2 + \left(\hat{y}_{t+1}^{(r)} - \hat{y}_{t+1}^{(u)}\right)^2 \right]
\end{equation}

I report Clark–West statistics for nested forecast comparisons in Table \ref{tab:cw_test}. The statistics are computed from the adjusted loss differential using heteroskedasticity- and autocorrelation-robust (Newey–West) standard errors appropriate for $h$-step-ahead forecasts. The alternative is one-sided (the augmented model is better). Across targets and horizons, the Clark and West t-statistics are positive and large, indicating that adding transcript-based textual features reduces forecast MSE relative to the benchmark. The implied mean-squared-error reductions are economically meaningful, typically about 6–13\%.

\paragraph{Is It Rational for Financial Analysts to Pay Attention to Corporate Communication?}

Table \ref{tab:improv_over_analysts1} compares the predictive accuracy of the proposed model to analysts' forecasts across different horizons, thereby assessing whether it is rational for financial analysts to pay attention to corporate communication. The Mean Squared Error (MSE) values represent the baseline forecast error, while the ``Gains'' columns indicate percentage improvements relative to this benchmark when incorporating textual and numerical information.

At the one-year horizon, the model achieves only a modest improvement over analysts' forecasts, with a total gain of 8.38\%. This limited enhancement suggests that the immediate informational content of corporate communication may be either already incorporated into analysts' assessments or overshadowed by short-term quantitative signals. It is also possible that analysts initially overreact to qualitative cues, leading to a temporary reduction in the predictive value of textual information.

By contrast, the two- and three-year horizons exhibit substantial gains of 27.50\% and 31.64\%, respectively, indicating that the incorporation of textual variables meaningfully enhances forecast accuracy over longer periods. This pattern implies that the qualitative dimensions of managerial discourse, such as tone, strategic emphasis, and forward-looking statements, capture aspects of firms' trajectories that unfold only gradually and are not fully reflected in contemporaneous numerical data.

Overall, the results suggest that while short-term expectations may be adequately informed by quantitative fundamentals, the integration of textual information becomes increasingly valuable for medium- to long-term forecasting. Hence, paying attention to corporate communication is rational for financial analysts seeking to understand firms'

\subsection{Measuring the Impact of Fundamental News}

\begin{figure}
\centering
\caption{\textbf{Predicted Effects of Fundamental Information}}
\caption*{
This figure shows partial dependence plots (PDPs) of predicted 1-, 2-, 3-year outcomes with respect to Fundamental Information, proxied by Standardized Unexpected Earnings (SUE). PDPs show how the model's predictions vary with a given feature while averaging over the empirical distribution 
of all other features.  Panels show, respectively: (top left) \textit{Analysts' Expected Change in Earnings} (analysts' consensus after the earnings call minus the value just realized), 
(right) \textit{Forecast Disagreement} (the standard deviation of analysts' consensus after the earnings call), and 
(bottom left) \textit{Realized Change in Earnings} (the value to be realized in one, two, or three years minus the value just realized).. The line traces the model's average prediction as SUE varies (holding other features at their empirical distribution).
}
\label{fig:pdp_sue}
\includegraphics[width=1.1\linewidth]{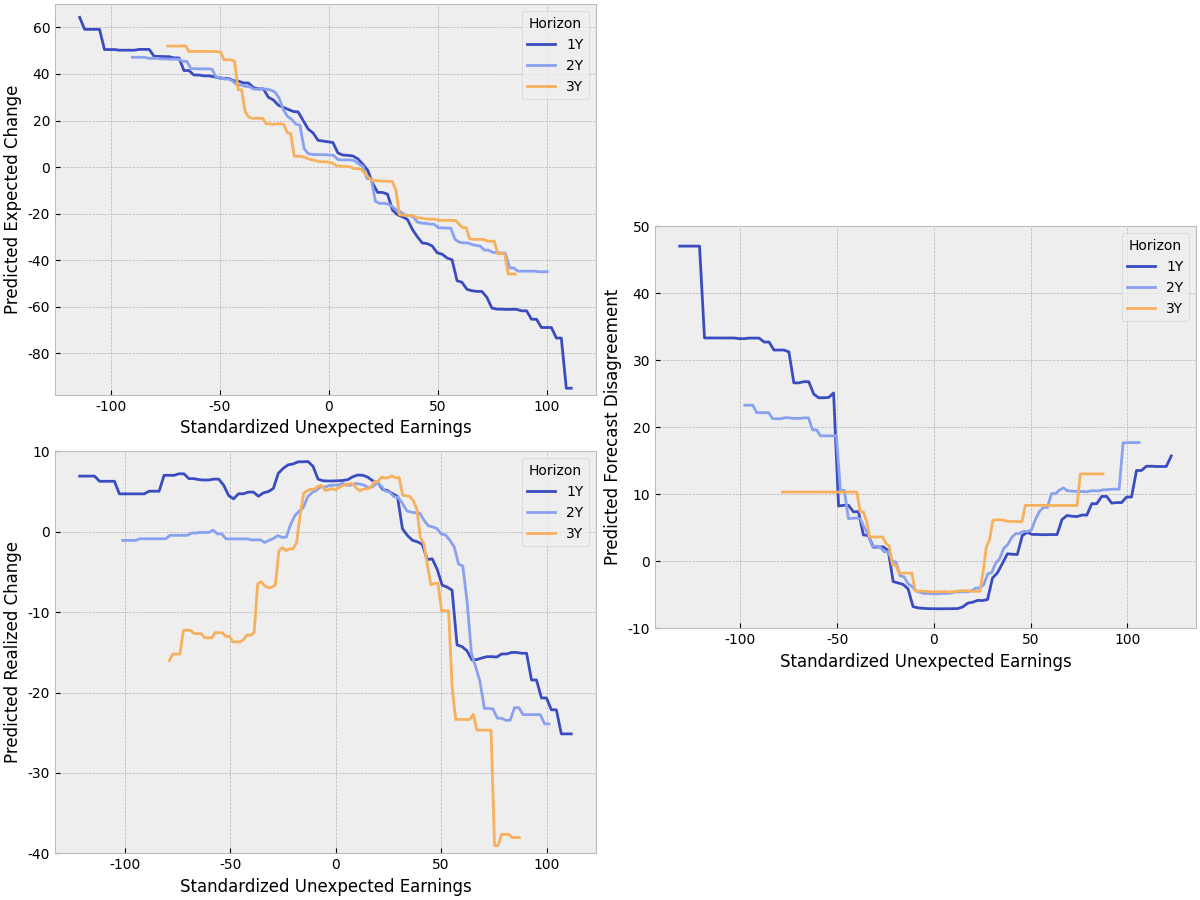}
\label{fig:fundamental_news}
\end{figure}

Understanding how analysts process fundamental news is central to evaluating the efficiency of their forecast revisions. Earnings announcements and related disclosures convey both quantitative and qualitative information about firms' underlying fundamentals, and the way analysts interpret these signals can shape the dynamics of market expectations. To assess this process, I measure the impact of fundamental news on analysts' forecast updates following earnings calls. \citet{Easterwood1999} show that analysts systematically underreact to bad news and overreact to good news, generating predictable forecast errors around earnings announcements. To relate my setting to this classic result—and to investigate how these dynamics extend to the context of earnings calls—I examine the relationship between standardized unexpected earnings (SUE) and analysts' forecasts. Figure \ref{fig:pdp_sue} reports partial dependence plots (PDPs) of the one-year targets on SUE. These PDP charts trace the marginal effect of a predictor on the model outcome, averaging over the distribution of other covariates, and thus provide a transparent view of how forecasts vary with earnings surprises. 

The patterns reveal familiar asymmetries in analysts' reactions: (a) forecast errors rise monotonically with SUE, with a steep transition around modest negative surprises and saturation in the tails; (b) realized changes in earnings exhibit an asymmetric, step-like increase after large negative surprises, plateauing and then slightly tapering for very positive surprises, consistent with mean reversion; (c) expected forecast changes decline with SUE---negative surprises are followed by upward revisions, while positive surprises are tempered by smaller or even negative adjustments; and (d) forecast disagreement is U-shaped, lowest near zero surprise and higher at the extremes, suggesting greater dispersion when news is unusually good or bad.  

I compute the inter-quartile shift, a "typical" move in SUE, (from the 25th to the 75th percentile) in 
standardized unexpected earnings (SUE) as the baseline variation. These results provide a benchmark against which, in later sections, I compare the 
effects of language morphing with large language models, thereby assessing whether 
earnings call framing can move forecasts in ways comparable to fundamental earnings news.

\subsection{Discussion}
\label{sec:discussion1}

The results provide clear evidence that financial analysts attend to earnings calls and that such attention has incremental informational value beyond traditional quantitative predictors. Across specifications, augmenting benchmark models based on accounting variables and stock characteristics with textual features from earnings calls improves predictive performance. While out-of-sample \(R^{2}\) remains a noisy metric, the Clark and West \citep{Clark2007} test consistently indicates that the incremental forecasting value of the narrative component is genuine rather than spurious.

These findings imply that analysts do not rely exclusively on hard data when forming expectations; they also process qualitative and contextual dimensions of managerial communication. From an informational-efficiency perspective, this is potentially beneficial: even models predicting realized earnings improve when textual features are included, suggesting that the narratives emphasized in earnings calls embed information about fundamentals that is not fully captured by standard predictors.

At the same time, the nature of analysts' attention raises important open questions. It remains to be established whether analysts focus on the same narrative dimensions that predict realized earnings—thereby incorporating value-relevant information in a manner consistent with rational processing—or whether attention is drawn to different, potentially more salient but less informative aspects of corporate communication. Disentangling these possibilities is essential to determine whether narratives primarily operate as carriers of genuine information or as channels for behavioral bias in financial markets.

\section{Counterfactual Earnings Calls with Large Language Models}
\label{sec:counterfactual_res}

The ideal experimental setting for isolating the causal effect of language on analyst perception would resemble a controlled environment in which a group of financial analysts resets their priors and issues new forecasts in response to alternative narrative formulations of the same earnings call. Such a laboratory-style manipulation is infeasible in practice. To address this challenge, I adopt a novel in-silico approach that leverages recent advances in natural language processing (NLP). While early NLP techniques enabled the transformation of textual data into vector representations suitable for econometric modeling, they lacked the capacity to simulate variations in rhetorical framing or tonal emphasis. The emergence of large language models (LLMs) overcomes this limitation by enabling the systematic generation of semantically equivalent, but stylistically distinct, versions of CEO earnings call presentations. This technological development allows for the construction of a counterfactual framework in which the informational content remains fixed, while linguistic framing is varied exogenously—thus permitting a thorough analysis of language effects on analyst behavior.

To do so, I generate synthetic variants of management speech using large language models (LLMs) implemented through the Hugging Face Transformers library. Specifically, I employ Meta's LLaMA 3 70B-Instruct model, a chat-optimized transformer fine-tuned for instruction-following and coherent text generation in conversational contexts.

For each document, I generate the output paragraph by paragraph in sequential forward passes, ensuring internal consistency and preserving global structure. To control the realism and interpretability of the generated texts, I apply prompt engineering techniques that guide the model to preserve numerical content (e.g., financial figures) while varying qualitative language. Specifically, I structure prompts in a way that keeps key values unchanged and induces linguistic transformations within a narrow range of word count variation, thereby enabling controlled comparisons between original and modified statements. See Section~\ref{section:prompts} in the appendix for further details.

This setup allows for systematic manipulation of the rhetorical framing and tone of the earnings call content, holding constant the quantitative core of the message.

In this regard, look-ahead bias does not affect our results because we train the predictive model ex ante, apply modifications through prompting to the LLM, and evaluate performance exclusively on out-of-sample morphed embeddings.

\subsection{Methodology for Language Generation}
Language generation is a high-dimensional statistical process that depends on the careful tuning of hyper-parameters governing the model's stochastic output. Unlike conventional econometric models, there are no universally accepted guidelines for tuning these parameters, as their effectiveness depends on complex interactions and the specific textual context. I adopt standard settings widely used in practice. The temperature parameter controls the entropy of the sampling distribution; lower values yield more deterministic output. I set \texttt{temperature} to $0.7$, a commonly used value in production settings such as ChatGPT. The \texttt{top p} parameter (set to $0.95$) constrains sampling to the smallest set of candidate tokens whose cumulative probability exceeds $95$\%, thereby excluding implausible continuations. These parameters jointly influence the diversity, coherence, and rhetorical variation of the generated text, and thus have meaningful implications for the empirical content  and the realism of simulated earnings calls.

\subsection{LLM as a Judge}

While large language models enable scalable counterfactual text generation, they remain susceptible to hallucinations that can undermine empirical validity. For example, a synthetically altered earnings call transcript might inadvertently report incorrect earnings figures, thereby biasing the predictive accuracy metrics reported in this paper. Even with carefully engineered prompts, instruction‐tuned chat variants can introduce spurious references to unrelated firms, stray numerical data, or extraneous statements drawn from their training corpus. Without an ex post validation mechanism, these unintended deviations threaten to produce misleading inferences and spurious results.

This unintended behavior, commonly referred to as ``hallucination'' in the ML literature, has been documented extensively by computer scientists and practitioners (e.g., \citealp{Huang2024}). Relying on manual inspection to verify whether prompt‐driven modifications have been executed correctly is both laborious and inherently subjective, exposing the analysis to evaluator bias. To overcome these limitations and ensure the internal validity of our counterfactual text experiments, I implement an automated ``judge'' using another instance of the model described above, Meta's LLaMA 3 70B-Instruct. LLM‑as‑a‑judge functions as an automated validation layer: a pretrained LLM assigns integer scores to both original and counterfactual transcripts along predefined dimensions—such as confidence, sentiment, and guidance strength—while flagging any spurious insertions (e.g., hallucinated figures or false statements). \citep{Li2024} 

By comparing these scores across versions, I ask the LLM to classify each morphed transcript 
into one of three categories:

\begin{enumerate}
\item \textbf{Yes} — the morphing was executed correctly and the language modification is clear and evident;
\item \textbf{I am unable to determine} — the morphing quality is uncertain;
\item \textbf{No} — the morphing is inadequate.
\end{enumerate}

I discard every variant that fails to meet the highest quality standard, thereby ensuring that 
the predictive analyses are based exclusively on precisely controlled linguistic change. The details on the prompt I use are in Section~\ref{section:judge-prompts}. 

This methodology helps create and validate credible textual counterfactuals by establishing an objective and reproducible mechanism for assessing the quality of generated texts. The LLM-as-a-judge framework acts as a consistency filter that ensures counterfactuals preserve the informational core of the original transcript while exhibiting a controlled and interpretable linguistic modification. By systematically scoring and classifying each counterfactual according to predefined semantic and stylistic dimensions, the judge model operationalizes the notion of ``credible'' counterfactuals—texts that differ meaningfully in the intended attribute (e.g., tone or sentiment) yet remain faithful to the underlying economic message.

In doing so, the approach mitigates two critical threats to internal validity. First, it minimizes the risk of \textit{semantic drift}, that is, the introduction of unintended changes to factual or contextual content that could distort the causal interpretation of downstream results. Second, it limits \textit{evaluation bias} by replacing subjective human judgment with an automated and transparent scoring procedure. Because the same pretrained model family is used for both generation and evaluation, the validation process benefits from consistent representational semantics across tasks, enhancing the reliability of the comparisons.

\subsection{Narratives and Counterfactual Earnings Calls}
\label{sec:narratives}

To isolate the effect of language on analyst forecasts, I generate counterfactual versions of earnings call transcripts in which the underlying quantitative content is held constant, but the linguistic framing is systematically modified.

The goal of this exercise is to assess how much of the analyst reaction can be attributed to style and delivery, rather than to substance. More in detail, I generate versions of the same earnings call that vary according to a number of linguistic characteristics, and evaluate how a classifier trained on historical data responds to each variant. The linguistic characteristics I use are:

\begin{itemize}
\item \textbf{Confidence}: Refers to how assertive and in-control the speaker sounds. This is adjusted by prompting the model to adopt the tone of a confident CEO, using decisive language, avoiding hedging, and conveying authority over strategy and performance. 

\citep{Mayew2012, DelaParra2024}

\item \textbf{Global Focus}: Captures the emphasis placed on broader macroeconomic and industry-wide conditions. The model is guided to highlight external forces, such as inflation, regulation, or global trends, and to link them explicitly to the firm's performance. 

\citep{Song2025, Link2023, Hassan2019}

\item \textbf{Guidance}: Reflects how specific and forward-looking the company's statements are. Stronger guidance is elicited by prompting clear, actionable expectations, directional indicators, and strategic intent, avoiding vague or generic phrasing. 

\citep{Call2024, Bozanic2018, Anilowski2007}

\item \textbf{Sentiment}: Denotes the overall emotional tone of the message. The text is modified to sound more positive and optimistic by emphasizing momentum, wins, and enthusiasm for the company's future. 

\citep{Price2012, Huang2014}

\item \textbf{Jargon}: Indicates the degree of technical or finance-specific language used. Higher jargon is introduced through prompts that incorporate institutional and domain-specific terminology without changing the underlying facts.

\citep{Li2008, Miller2010}

\item \textbf{Uncertainty}: Reflects the visibility and emphasis of risk-related language. To increase risk salience, prompts guide the model to explicitly highlight uncertainties, market challenges, and operational risks in a measured and professional tone. 

\citep{Kravet2013, Lyle2023}
\end{itemize}


Because the generated calls differ only in their language, any change in the model's prediction can be interpreted as an effect of linguistic framing on analyst perception. This setup allows me to quantify the persuasive power of narrative, assess the potential for strategic disclosure, and explore whether firms could influence analyst expectations through subtle shifts in communication—without altering the underlying message related to financial fundamentals.

The methodology also sheds light on the boundaries of predictability: how far a firm's rhetorical adjustments can shift market expectations, and whether such shifts align with or deviate from future fundamentals.

\begin{figure}
\captionsetup{labelfont={sc}}
\caption{\textbf{Geometry of Original and Morphed Earnings Calls}}
\caption*{This figure shows UMAP-computed \citep{McInnes2018} two-dimensional projections of text embeddings for original and morphed earnings call transcripts. Each panel is a 30×30 bins two-dimensional histogram; darker cells indicate higher local transcript density. The UMAP algorithm approximately preserves neighborhood structure, so transcripts with similar topics lie near one another. The last row shows the original (un-morphed) embeddings for comparison, and the 2D coordinates are to be interpreted in the comparison across morphing directions, so corresponding regions tend to represent the same topics across embeddings.}
\label{fig:umap_mapping}
\centering
\includegraphics[width=\textwidth]{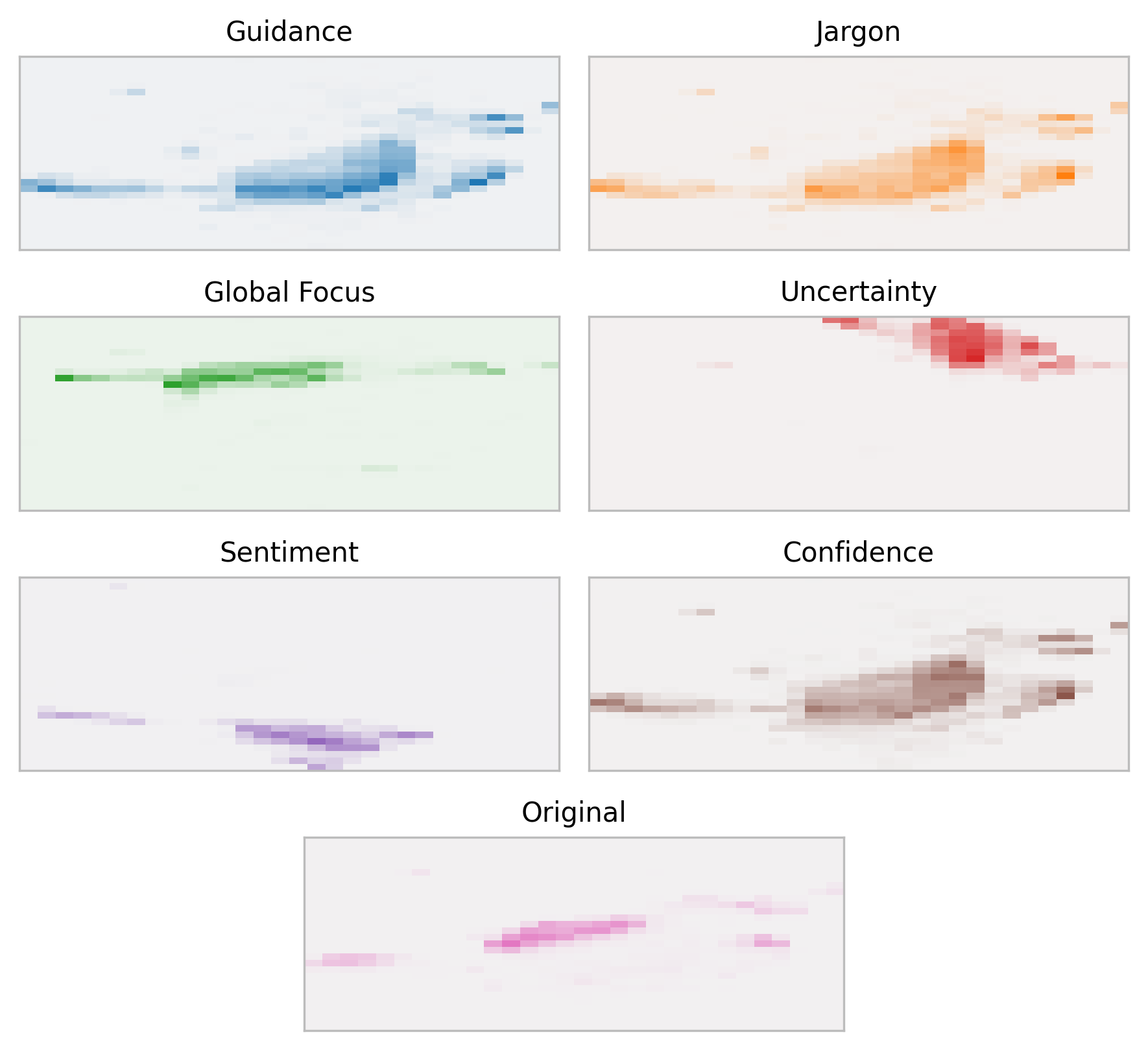}
\end{figure}

I illustrate this morphing exercise in Figure~\ref{fig:umap_mapping}, where I project both the 
original and morphed transcripts into a two-dimensional semantic space. Each panel 
corresponds to a distinct linguistic attribute: guidance strength, jargon density, speaker 
confidence, global focus, sentiment, and risks mentioned,  
the original baseline in the bottom panel. The heatmaps reveal that morphing pushes transcripts toward the extremes 
of these attributes, generating clearly differentiated clusters. This demonstrates that the 
procedure succeeds in producing systematically varied earnings calls, thereby enabling the 
in-silico experimental design.

To further illustrate the results of my morphing exercise, I report Table \ref{tab:morphed_examples}, which illustrates how the original text of an earnings call can be systematically morphed along distinct narrative dimensions. Each row presents a representative passage from an original managerial statement alongside a transformed version that amplifies a specific linguistic or narrative feature. The six dimension i.e., Guidance, Jargon, Confidence, Global Focus, Sentiment, and Uncertainty, capture complementary aspects of managerial communication that may influence analysts' interpretation of firm fundamentals. These examples serve to clarify the nature of the textual transformations applied in the empirical analysis and to make the corresponding treatments intuitively interpretable.

\paragraph{Relevance of Linguistic Dimensions.}
Although these narratives \citep{Shiller2017} draw upon established concepts in economics and finance, their direct connection to analyst behavior remains largely unexplored. In this paragraph, I examine whether the six previously defined linguistic ``factors'' are predictive of analysts' expected changes in earnings, forecast dispersion, and realized earnings changes.

\begin{figure}
\centering
\caption{\textbf{Term Structure of Textual Predictive Power}}
\caption*{
This chart reports the out-of-sample explanatory power, expressed in R-squared, of a model based solely on textual features from earnings calls across different forecast horizons (1-year ahead, 2-year ahead, and 3-year ahead) for three outcomes: (i) \textit{Analysts' Expected Change in Earnings} (analysts' consensus after the earnings call minus the value just realized), 
(ii) \textit{Forecast Disagreement} (the standard deviation of analysts' consensus after the earnings call), and 
(iii) \textit{Realized Change in Earnings} (the value to be realized in one, two, or three years minus the value just realized).
}
\includegraphics[width=\textwidth]{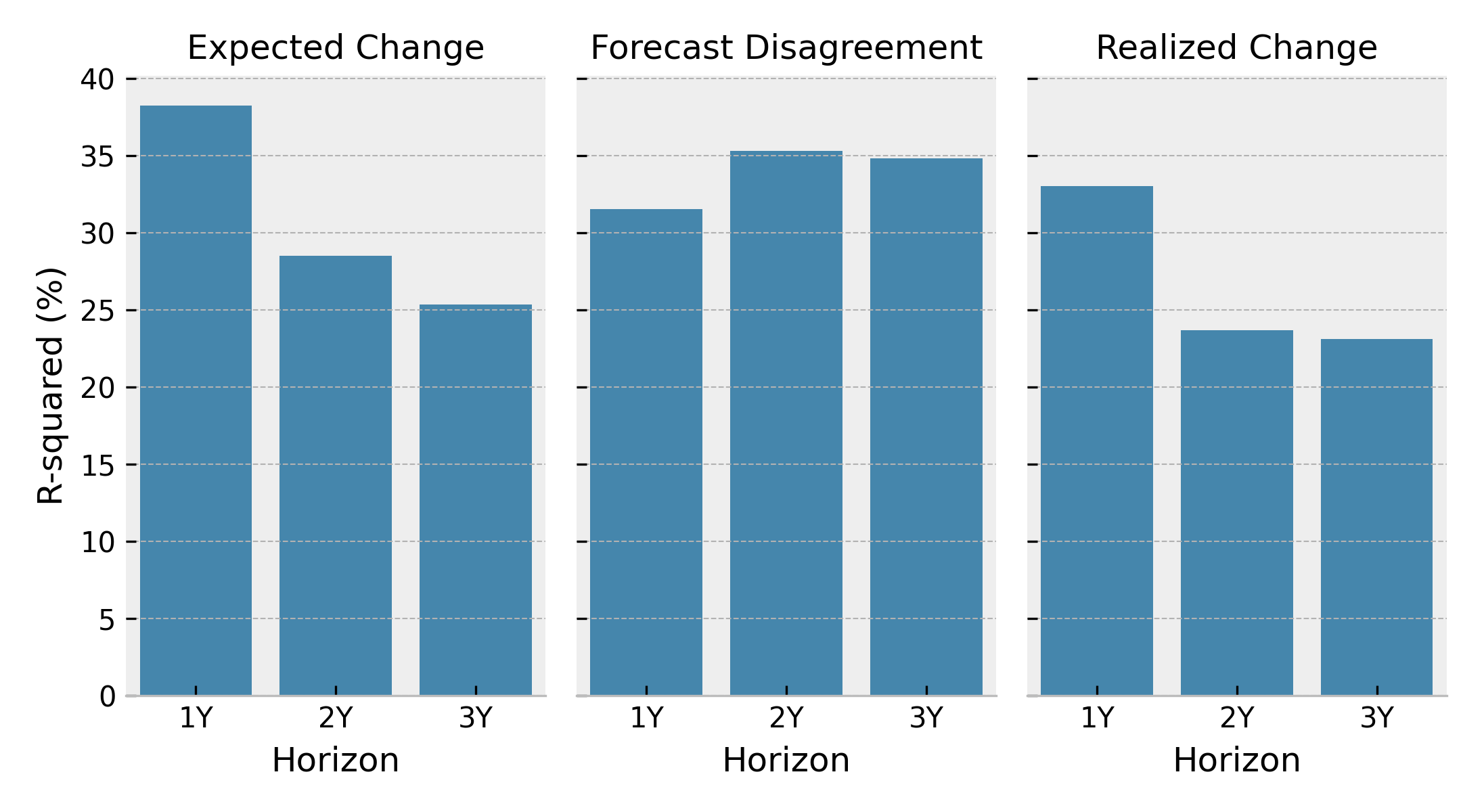}
\label{fig:just text}
\end{figure}

The analysis follows the estimation framework underlying Figure \ref{fig:just text}, with the key difference that, instead of employing the full 768-dimensional textual embeddings, I use six-dimensional vectors that capture the intensity of each specific linguistic dimension within the earnings call. To measure these dimensions, I employ Llama 3.1, prompting each chunk of the transcript to rate the presence of the corresponding linguistic factor on a scale from 1 to 10. I then compute the average score across all chunks for each call.

\begin{figure}
\centering
\caption{\textbf{Term structure of Linguistic Factors Predictive Power}}
\caption*{
This chart reports the out-of-sample predictive power, expressed in R-squared, that incorporate only six linguistic factors quantifying the presence of key narratives in earnings call transcripts, specifically Guidance, Jargon, Confidence, Global Focus, Sentiment, and Uncertainty.
The predictive results are shown for different forecast horizons (1-year ahead, 2-year ahead, and 3-year ahead) and for three outcomes: (i) \textit{Analysts' Expected Change in Earnings} (analysts' consensus after the earnings call minus the value just realized), 
(ii) \textit{Forecast Disagreement} (the standard deviation of analysts' consensus after the earnings call), and 
(iii) \textit{Realized Change in Earnings} (the value to be realized in one, two, or three years minus the value just realized).
}
\label{fig:just text factors}
\includegraphics[width=\textwidth]{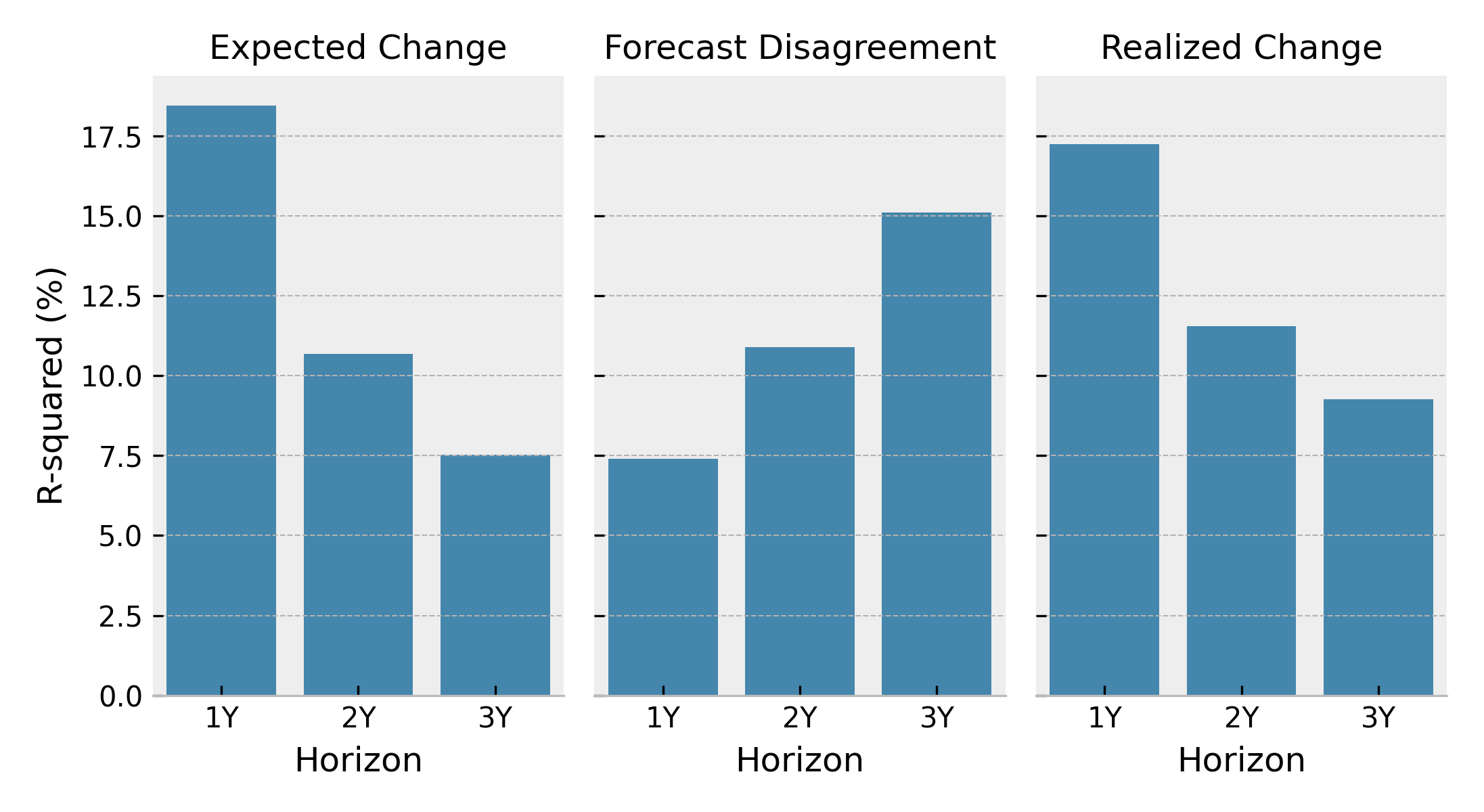}
\end{figure}

As shown in Figure \ref{fig:just text factors}, these linguistic factors exhibit substantial predictive power for analyst expectations and realized earnings changes. They account for approximately 25–35\% of the explanatory capacity of the full textual embeddings, as reported in Figure \ref{fig:just text}, suggesting that these narrative dimensions capture core mechanisms through which language in earnings calls shapes analysts' beliefs and firm performance.

\subsection{Predicted Treatment Effects (PTEs) induced by Narratives}

To summarize, I use one of the most recent language models, \emph{Llama 3}, to generate morphed versions of each earnings call transcript, specifically designed to emphasize and incorporate a given narrative dimension. Formally, the transformation can be expressed as
\begin{equation}
\mathbf{LLM(}\text{EC}_{i,t}) = \text{Morphed EC}_{i,t},
\end{equation}

for each company $i$ holding an earnings call at $t$, where the large language model produces a modified embedding that reflects the targeted linguistic theme. I then compute the \emph{Predicted Text Effect (PTE)} associated with the morphing procedure as
\begin{equation}
\text{PTE}_{i,t,morph} = F(\text{SC}_{i,t}, \text{Morphed EC}_{i,t}) - F(\text{SC}_{i,t}, \text{EC}_{i,t}),
\end{equation}
where $F(\text{fundamentals}, \text{textual features})$ is the model trained on the stock characteristics and the original earnings calls- The PTE measures the change in the model's predicted outcome when the textual representation is altered to highlight the specified theme, holding all other market-based firm-level and accounting predictors (\(\text{SC}_{i,t}\)) constant.

I use the \textit{in-silico} laboratory to compute the predicted treatment effect (PTE) of applying linguistic morphings to the management presentations in earnings calls, while holding all other factors constant. The laboratory setting allows me to systematically vary tone or topical emphasis in the transcripts and then trace the implied change in analysts' reactions, as captured by the machine-learning models described in the previous sections.  


The PTE therefore represents the marginal effect of a counterfactual change in communication style, abstracting from any concurrent shifts in fundamentals or other explanatory variables, as I use the model specification including all the available features. 

\begin{figure}[htbp]
\centering
\captionsetup{labelfont={sc}}
\caption{\textbf{Narratives in Analysts Beliefs and Realized Earnings}}
\caption*{
This chart reports the average predicted treatment effects (PTEs) of the narratives (described in \hyperref[sec:counterfactual_res]{Section~\ref*{sec:counterfactual_res}}) on \textit{Analysts' Expected Change in Earnings} (analysts' consensus after the earnings call minus the value just realized) and \textit{Realized Change in Earnings} (the value to be realized in one, two, or three years minus the value just realized). PTEs are computed as the difference between the output of the model trained on fundamentals and textual features ($F(\text{fundamentals}, \text{text})$) when earnings call transcripts are morphed to emphasize a given narrative and its output when evaluated on the original transcripts. \textit{Fundamental News} corresponds to the effect associated with an inter-quartile change in the empirical distribution of Standardized Unexpected Earnings (SUE), as predicted by the same model, $F(\text{fundamentals}, \text{text})$. The \textit{Difference} bar reports the gap between the \textit{Analysts' Expected Change in Earnings} and \textit{Realized Change in Earnings}
}
\label{fig:treat1}
\includegraphics[width=0.9\textwidth]{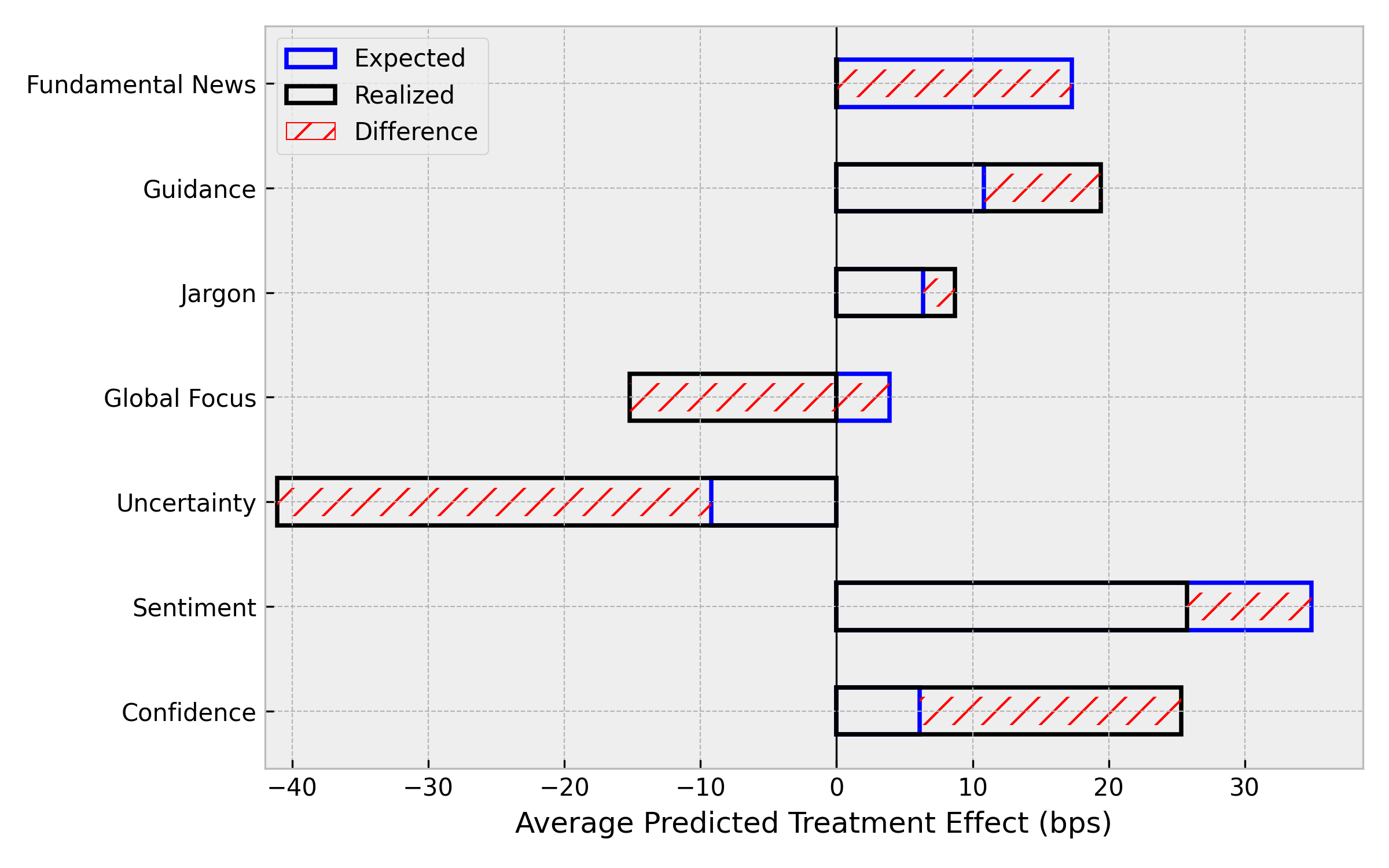}
\end{figure}

\begin{figure}[htbp]
\centering
\caption{\textbf{Narratives and Forecast Disagreement}}
\caption*{
This chart reports the average predicted treatment effects (PTEs) of the narratives (described in \hyperref[sec:counterfactual_res]{Section~\ref*{sec:counterfactual_res}}) on \textit{Forecast Disagreement} (the standard deviation of analysts' consensus after the earnings call). PTEs are computed as the difference between the output of the model trained on fundamentals and textual features ($F(\text{fundamentals}, \text{text})$) when earnings call transcripts are morphed to emphasize a given narrative and its output when evaluated on the original transcripts. \textit{Fundamental News} corresponds to the effect associated with an inter-quartile change in the empirical distribution of Standardized Unexpected Earnings (SUE), as predicted by the same model, $F(\text{fundamentals}, \text{text})$.
}
\label{fig:treat2}
\includegraphics[width=0.9\textwidth]{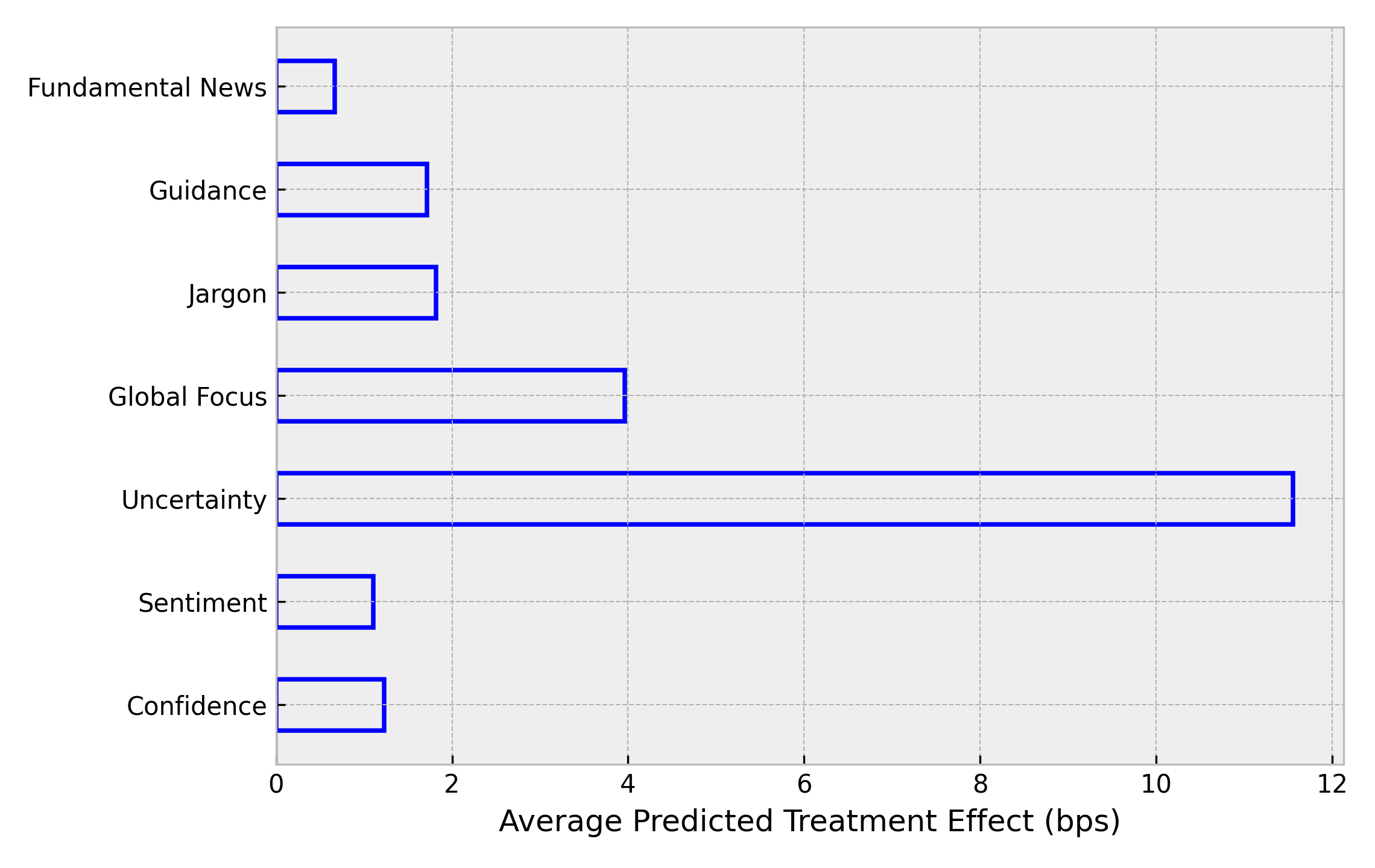}
\end{figure}

Figures \ref{fig:treat1} and \ref{fig:treat2} present the average predicted treatment effects (PTEs), expressed in basis points, for each linguistic morph applied to earnings calls. These effects quantify how shifts in communication style—independent of any change in firm fundamentals—alter analysts' forecasts and realized outcomes. To benchmark the magnitude of these effects, I also report the model-implied response to an inter-quartile change in fundamental earnings news, defined as the difference in predicted outcomes when standardized unexpected earnings (SUE) move from the 25th to the 75th percentile of its distribution, as computed in prior sections.

Figure~\ref{fig:treat1} displays the average predicted treatment effects for both expected and realized outcomes across linguistic dimensions. The chart highlights that narrative tone and content meaningfully shape the information that analysts extract from corporate communication. The most pronounced effects emerge in the ``Sentiment'' and ``Confidence'' dimensions, where both expected and realized responses are positive and of comparable magnitude, indicating that optimistic or assertive managerial language systematically elevates analysts' forecasts and is partially validated by subsequent firm performance. Conversely, ``Uncertainty'' exerts a strong negative effect on realized outcomes, suggesting that ambiguous or risk-laden discourse dampens analysts' expectations and is indeed associated with weaker future fundamentals. The ``Global Focus'' and ``Guidance'' categories exhibit moderate positive effects, consistent with the notion that outward-looking and directive communication enhances perceived managerial clarity. Overall, the close alignment between expected and realized treatment effects supports the model's interpretability, as it suggests that linguistic features influence beliefs in economically meaningful ways.

Figure~\ref{fig:treat2} isolates the predicted treatment effects of each narrative dimension on \textit{forecast disagreement}, capturing how language shapes the dispersion of analysts' expectations rather than their central tendency. Here, ``Uncertainty'' again stands out as the dominant driver, with a markedly higher treatment effect than any other linguistic feature. This finding aligns with the theoretical expectation that ambiguous communication amplifies interpretive heterogeneity among analysts, leading to greater disagreement. However, confident, sentiment-rich, and globally oriented discourse also increase forecast dispersion, albeit to a lesser extent. This pattern suggests that even positive or outward-looking narratives can introduce divergent interpretations among analysts, possibly because such language allows multiple readings of firms' strategic intent or the macroeconomic implications of managerial optimism. Together, Figures~\ref{fig:treat1} and~\ref{fig:treat2} underscore that managerial narratives not only shift the level of analysts' forecasts but also shape the distribution of beliefs across forecasters, revealing a dual channel through which corporate communication affects market expectations.

\subsection{Discussion}
\label{sec:discussion2}

The finding that language---often conveying no new quantitative disclosure and reflecting managerial tone rather than substantive information---can nonetheless induce sizeable revisions in analysts' forecasts underscores the powerful role of corporate communication in shaping market expectations. This result aligns with the broader view, emphasized by \citet{Shiller2017}, that narratives act as vehicles for economic meaning: they organize complex information into coherent stories that influence behavior even when their quantitative content is limited. Beyond this, the analysis shows that textual features extracted from earnings calls provide incremental predictive power over traditional accounting- and market-based predictors. The soft information embedded in managerial speech thus captures dimensions of analysts' responses not explained by fundamentals alone. In this sense, earnings calls illustrate how the transmission of information in financial markets depends not only on the release of hard data, but also on framing, emphasis, and rhetorical style.

At the same time, the results highlight a deeper tension in belief formation. Financial analysts absorb and process a vast amount of quantitative and qualitative information, yet what ultimately drives realized earnings is largely tied to underlying risk and its evolution over time. Whether analysts attend to the same narratives that carry genuine information about these risks---or instead to stories that merely resonate with prevailing sentiment---remains an open question. Distinguishing between these two forms of narrative attention is essential for understanding whether the influence of corporate communication reflects rational information processing or the behavioral power of persuasion in financial markets.

\section{Conclusion}
\label{sec:conclus}

Narratives are ubiquitous in financial markets. I show that, in the context of corporate earnings calls, they affect analysts' expectations: financial analysts rationally incorporate narrative information, which also proves useful in predicting future realized changes in earnings.

To investigate how financial analysts under- or over-react to specific narratives, I propose a novel framework that performs human-like edits (morphs) to the text of earnings calls, altering the predominant narrative within each document.

I then input the morphed transcripts into the machine learning models trained on analysts' expectations and realized earnings. The difference between the new predicted outcome and the original prediction represents the predicted treatment effect of the specific narrative.

This methodology allows me to precisely compute analysts' over- and under-reaction to a given narrative. I find that financial analysts severely under-react to the narrative of uncertainty and over-react to the optimistic narrative and to global-focus framing. I also find that disagreement among analysts is mainly driven by the narrative of uncertainty and global-focus perspective.

The framework developed in this paper offers a practical and scalable tool for market participants. For corporate managers, it provides a means to strategically design corporate communication by experimenting, in an \textit{in-silico} environment, with alternative framings of the same message to anticipate potential market reactions. For analysts, it highlights the importance of distinguishing substantive information from rhetorical framing and of exercising caution when firms emphasize certain topics for strategic rather than informational reasons. For researchers, it opens new avenues for examining the role of narratives in financial markets and offers fresh insights into the mechanisms underlying belief formation among financial analysts.

This approach can be extended to other domains of corporate disclosure, such as sustainability reporting, M\&A announcements, or regulatory filings, where language plays a central role in shaping investor beliefs. It can likewise be applied to macroeconomic and policy communication—such as FOMC press conferences or central bank statements—to study how variations in tone, emphasis, or framing influence market reactions. More broadly, the paper illustrates how advances in natural language processing can be leveraged to address long-standing questions in financial economics concerning the interaction between information and beliefs.

\clearpage

\section{Figures}

\clearpage

\clearpage

\begin{figure}[ht]
\caption{\textbf{Timeline of the Earnings Announcement Cycle}}
\caption*{This figure illustrates the standard sequence of events in the corporate earnings announcement cycle. At time $t$, the firm reports realized earnings for the preceding period ($Y_{t-1}$) and files its Form 10-K with the SEC, which includes the income statement, balance sheet, and statement of cash flows. Around the same time, management holds an earnings conference call with investors and financial analysts to discuss recent performance and provide forward-looking guidance. Analysts then revise their earnings forecasts for the current period ($Y_t$) in light of the new information. At time $t + h$, where $h$ denotes the forecast horizon, the firm reports the realized earnings for the forecasted period ($Y_{t+h-1}$).}
\label{fig:timeline}
\begin{center}
\begin{tikzpicture}[
node distance=1.2cm,
every node/.style={font=\small},
arrow/.style={-{Latex[length=3mm]}, thick},
event/.style={text width=6cm, align=left, rounded corners, draw=black, fill=gray!10, inner sep=12pt}
]
\node[event] (t) {\textbf{Time $t$}
\begin{enumerate}
\item The company discloses realized earnings for the previous period (Y$_{t-1}$) and files its 10-K report with the SEC.
\item A corporate earnings call is held where management delivers prepared remarks to financial analysts and investors.
\item Financial analysts review the latest information and revise their earnings forecasts for the current year (Y$_{t}$).
\end{enumerate}
};

\node[event, right=of t] (t2) {\textbf{Time $t+h$}
\begin{itemize}
    \item The company discloses the realized earnings for the forecasted period (Y$_{t+h-1}$).
\end{itemize}
};

\draw[arrow] (t) -- (t2);
\end{tikzpicture}
\end{center}
\end{figure}

\clearpage


\clearpage

\clearpage

\clearpage

\clearpage

\newcommand{\ptescale}{0.3}

\clearpage

\clearpage
\section{Tables}

\begin{table}[ht]
\centering
\caption{\textbf{Distribution of Sample Observations Across Forecast Horizons}}
\caption*{This table reports, for the one-, two-, and three-year ahead forecast horizons, the number of observations, their percentage share of the total sample, the number of companies, the number of brokers, and the number of analysts employed by those brokers. The sample covers the period 2006–2023.}
\label{tab:fpi_counts}
\begin{tabular}{lrrrrrr}
\toprule
\rowcolor{gray!20} Horizon & \# Obs & \% Obs & \# Companies & \# Brokers & \# Analysts \\
\midrule
1Y & 34798 & 45 & 5338 & 611 & 8242 \\
2Y & 27808 & 36 & 4654 & 563 & 7388 \\
3Y & 15234 & 19 & 3240 & 367 & 3662 \\
\bottomrule
\end{tabular}
\end{table}
\clearpage

\begin{table}[ht]
\centering
\caption{\textbf{Summary Statistics for Target Variables}}
\caption*{This table reports summary statistics for the three target variables used in the forecasting exercises. 
The variables are: (i) \textit{Analysts' Expected Change in Earnings} (analysts' consensus after the earnings call minus the value just realized), 
(ii) \textit{Forecast Disagreement} (the standard deviation of analysts' consensus after the earnings call), and 
(iii) \textit{Realized Change in Earnings} (the value to be realized in one, two, or three years minus the value just realized). 
All variables are trimmed at the 5\% and 95\% levels. 
For clarity and comparability, all quantities are reported in basis points.}

\label{tab:summ-stats}
\scalebox{0.95}{
\begin{tabular}{lrrrrrrrr}
\multicolumn{9}{l}{\textbf{(A) Expected Change in Earnings}}\\
\multicolumn{9}{c}{}\\
\toprule
\rowcolor{gray!20} Horizon & Count & Mean & Std & Min & p25 & p50 & p75 & Max \\
\midrule
1Y & 31318 & 344.95 & 376.83 & -1227.68 & 221.45 & 410.90 & 568.75 & 1029.61 \\
2Y & 25028 & 517.20 & 303.90 & -565.25 & 361.98 & 528.82 & 696.09 & 1270.89 \\
3Y & 13714 & 619.48 & 304.38 & -227.74 & 433.20 & 607.21 & 798.32 & 1465.52 \\
\bottomrule
\multicolumn{9}{c}{}\\
\multicolumn{9}{l}{\textbf{(B) Forecast Disagreement}}\\
\multicolumn{9}{c}{}\\
\toprule
\rowcolor{gray!20} Horizon & Count & Mean & Std & Min & p25 & p50 & p75 & Max \\
\midrule
1Y & 29769 & 81.89 & 101.60 & 4.38 & 17.45 & 40.69 & 101.47 & 547.08 \\
2Y & 23491 & 103.83 & 114.33 & 9.86 & 29.56 & 58.81 & 130.34 & 615.78 \\
3Y & 9648 & 128.64 & 146.76 & 8.85 & 34.43 & 70.45 & 159.90 & 775.18 \\
\bottomrule
\multicolumn{9}{c}{}\\
\multicolumn{9}{l}{\textbf{(D) Realized Change in Earnings}}\\
\multicolumn{9}{c}{}\\
\toprule
\rowcolor{gray!20} Horizon & Count & Mean & Std & Min & p25 & p50 & p75 & Max \\
\midrule
1Y & 31316 & 264.91 & 500.37 & -1716.17 & 124.27 & 376.07 & 556.79 & 1152.66 \\
2Y & 25017 & 363.07 & 476.75 & -1228.29 & 148.23 & 422.05 & 646.46 & 1470.19 \\
3Y & 13707 & 439.19 & 479.46 & -1005.35 & 177.69 & 469.70 & 724.51 & 1662.47 \\
\bottomrule
\end{tabular}
}
\end{table}

\begin{table}[ht]
\centering
\caption{\textbf{Predictive Accuracy for Analyst Behavior and Earnings Outcomes}}
\caption*{This table reports the out-of-sample performance of predictive models expressed in terms of R-squared. 
I compare specifications using traditional features (market-based information and financial statement variables) with specifications augmented by textual embeddings from earnings call transcripts. 
The three target variables are: (i) \textit{Analysts' Expected Change in Earnings} (analysts' consensus after the earnings call minus the value just realized), (ii) \textit{Forecast Disagreement} (the standard deviation of analysts' consensus after the earnings call), and (iii) \textit{Realized Change in Earnings} (the value to be realized in one, two, or three years minus the value just realized). The table contains three columns: \textbf{R-squared (Fundamentals)}, reporting the explanatory power of the model using only stock characteristics and financial statement variables; \textbf{R-squared (Fundamentals + Text)}, reporting the explanatory power once textual embeddings from earnings call transcripts are added; and \textbf{Gain (\%)}, showing the percentage increase in R-squared attributable to the textual features.
}

\label{tab:r2_levels}
\scalebox{0.9}{
\begin{tabular}{lrrr}
\multicolumn{3}{l}{\textbf{(i) Expected Change in Earnings}} \\
\multicolumn{3}{l}{\textbf{}} \\
\toprule
\rowcolor{gray!20} Horizon & R-squared (Fundamentals) & R-squared (Fundamentals + Text) & Gain (\%) \\
\midrule
1Y & 68.45 & 69.35 & 1.32 \\
2Y & 56.38 & 57.07 & 1.22 \\
3Y & 49.43 & 50.15 & 1.46 \\
\bottomrule
\multicolumn{3}{l}{\textbf{}} \\
\multicolumn{3}{l}{\textbf{(ii) Forecast Disagreement}} \\
\multicolumn{3}{l}{\textbf{}} \\
\toprule
\rowcolor{gray!20} Horizon & R-squared (Fundamentals) & R-squared (Fundamentals + Text) & Gain (\%) \\
\midrule
1Y & 57.91 & 58.55 & 1.11 \\
2Y & 57.62 & 59.12 & 2.60 \\
3Y & 51.19 & 52.74 & 3.05 \\
\bottomrule
\multicolumn{3}{l}{\textbf{}} \\
\multicolumn{3}{l}{\textbf{(iii) Realized Change in Earnings}} \\
\multicolumn{3}{l}{\textbf{}} \\
\toprule
\rowcolor{gray!20} Horizon & R-squared (Fundamentals) & R-squared (Fundamentals + Text) & Gain (\%) \\
\midrule
1Y & 54.59 & 55.81 & 2.23 \\
2Y & 38.02 & 38.80 & 2.06 \\
3Y & 33.26 & 35.35 & 6.28 \\
\bottomrule
\end{tabular}
}
\end{table}

\clearpage


\begin{table}[ht!]
\centering
\caption{\textbf{Performance of the ML Model Relative to Analyst Forecasts}}
\caption*{This table compares the out-of-sample performance of the machine-learning model in predicting realized earnings at one-, two-, and three-year horizons with the accuracy of analyst forecasts. 
The Mean Squared Error (MSE) in the first row corresponds to the baseline forecast error implied by analyst predictions. 
Two specifications of the ML model are considered: a fundamentals-only model that includes stock characteristics and financial statement variables, and an expanded model that additionally incorporates textual features. 
\textbf{Fundamentals Gain} reports the percentage improvement delivered by the fundamentals-only model, while \textbf{Total Gain} reflects the combined improvement from fundamentals and textual features. 
Positive values indicate superior predictive accuracy relative to analyst forecasts.}
\label{tab:improv_over_analysts1}
\begin{tabular}{lrrr}
\toprule
\rowcolor{gray!20} Horizon & 1Y & 2Y & 3Y \\
\midrule
MSE (Forecast) & 0.001116 & 0.001784 & 0.002042 \\
Fundamentals Gain (\%) & 5.75 & 26.42 & 29.59 \\
Total Gain (\%) & 8.38 & 27.50 & 31.64 \\
\bottomrule
\end{tabular}
\end{table}

\clearpage

\begin{table}[ht]
\centering
\caption{\textbf{Improvement in Forecast Accuracy from Textual Features}}
\caption*{This table reports the results of the \citet{Clark2007} nested forecast comparison test assessing whether textual measures provide statistically significant out-of-sample gains when predicting both fundamentals and analyst behavior at one-, two-, and three-year forecast horizons. The three target variables are: (i) \textit{Analysts' Expected Change in Earnings} (analysts' consensus after the earnings call minus the value just realized), (ii) \textit{Forecast Disagreement} (the standard deviation of analysts' consensus after the earnings call), and (iii) \textit{Realized Change in Earnings} (the value to be realized in one, two, or three years minus the value just realized). The comparison involves two specifications: a benchmark model including stock characteristics and financial-statement variables, and an expanded model that additionally incorporates textual features. The one-sided alternative hypothesis tests whether the text-augmented model delivers superior forecasting performance relative to the benchmark. \textbf{Mean Squared Error (MSE) Reduction due to Text} reports the percentage decrease in test-set MSE attributable to textual information (adjusted for noise induced by model expansion), and \textbf{C\&W t-stat} provides the corresponding test statistic.}

\label{tab:cw_test}
\begin{tabular}{lrr}
\multicolumn{3}{l}{\textbf{(i) Expected Change in Earnings}} \\
\multicolumn{3}{l}{\textbf{}} \\
\toprule
\rowcolor{gray!20} Horizon & MSE Reduction due to Text (\%) & C\&W t-stat \\
\midrule
1Y & 9.70 & 11.67 \\
2Y & 12.13 & 11.06 \\
3Y & 8.10 & 7.32 \\
\bottomrule
\multicolumn{3}{l}{\textbf{}} \\
\multicolumn{3}{l}{\textbf{(ii) Forecast Disagreement}} \\
\multicolumn{3}{l}{\textbf{}} \\
\toprule
\rowcolor{gray!20} Horizon & MSE Reduction due to Text (\%) & C\&W t-stat \\
\midrule
1Y & 9.11 & 10.50 \\
2Y & 10.82 & 9.89 \\
3Y & 12.55 & 7.64 \\
\hline
\multicolumn{3}{l}{\textbf{}} \\
\multicolumn{3}{l}{\textbf{(iii) Realized Change in Earnings}} \\
\multicolumn{3}{l}{\textbf{}} \\
\toprule
\rowcolor{gray!20} Horizon & MSE Reduction due to Text (\%) & C\&W t-stat \\
\midrule
1Y & 6.97 & 10.30 \\
2Y & 9.21 & 11.52 \\
3Y & 8.93 & 8.51 \\
\bottomrule
\end{tabular}
\end{table}

\begin{table}[htbp]
\centering
\caption{\textbf{Examples of Original and \textit{Morphed} Earnings Call Excerpts by Narrative}}
\caption*{This table provides illustrative pairs of excerpts corresponding to the \textit{morphing} exercise detailed in \hyperref[sec:counterfactual_res]{Section~\ref*{sec:counterfactual_res}}. The first column contains the passages from the original earnings call, while the second column reports the \textit{morphed} versions produced by the LLM when prompted to intensify a specific narrative.}
\label{tab:morphed_examples}
\renewcommand{\arraystretch}{1.3}
\scalebox{0.9}{
\begin{tabular}{@{}p{3cm}p{6.5cm}p{6.5cm}@{}}
\toprule
\textbf{Narrative} & \textbf{Original Excerpt} & \textbf{Morphed Excerpt} \\ 
\midrule

\textbf{Guidance} &
Looking ahead, we expect steady demand in our core markets and continued margin discipline. While some uncertainty remains, we're confident in our long-term strategy. &
Looking ahead, we anticipate revenue growth next quarter, driven by accelerating demand in our core markets. Operating margins should expand as we execute our cost initiatives. \\

\textbf{Jargon} &
We have streamlined operations to better align production with customer demand. &
We've optimized our end-to-end operational footprint through agile capacity realignment to ensure cross-functional synergies across key demand verticals. \\

\textbf{Global Focus} &
Our Q2 results were primarily shaped by the success of our new retail partnerships. &
Our Q2 results reflected broader macro dynamics--particularly the uptick in consumer confidence and stabilizing global supply chains--which supported retail expansion across regions. \\

\textbf{Uncertainty} &
We are encouraged by our progress and expect continued momentum next quarter. &
While we are encouraged by our progress, we remain cautious about potential headwinds--including raw material inflation, geopolitical tensions, and slower--than-expected demand recovery in Europe. \\

\textbf{Sentiment} &
This quarter's performance was in line with our expectations. &
This quarter's performance was excellent, surpassing expectations and reaffirming the strength of our business model. \\

\textbf{Confidence} &
We think our new product line will perform well, though it is still early in the rollout. &
Our new product line is already exceeding expectations, and we are absolutely confident it will continue to drive strong performance as adoption scales. \\

\bottomrule
\end{tabular}
}
\end{table}

\clearpage
\bibliography{bibliography} 
\clearpage
\appendix
\section*{Appendix}



\section{Identification of the latest version of earnings call transcripts}
\label{section:latest_version_transcripts}

I obtain the full text and associated metadata for earnings call events from the Capital IQ Transcripts database. Capital IQ Transcripts contains the full text of approximately 8{,}000 public-company events, including earnings calls, M\&A calls, and shareholder meetings. From this database, I extract all available earnings-call transcripts.

Each transcript undergoes four stages of editorial review: \textit{Preliminary} (real-time cleanup of voice-to-text with basic spellcheck), \textit{Edited} (accuracy pass with added metadata), \textit{Proofing} (external review for errors/classifications), and \textit{Audited} (final quality assurance).

Different versions are indicated by \texttt{transcriptCollectionTypeId}. Each version has its own \texttt{transcriptId}, and metadata fields—such as the event date or title—may exhibit minor revisions across versions even when they refer to the same underlying event. When multiple versions are available, I consider the most recent timestamp as typically the most accurate.

Because each event is indexed by \texttt{keydevid} and may pertain to multiple companies, I construct the dataset at the (\texttt{keydevid}, company) level. For each event–company pair, I retain the most up-to-date transcript—operationalized as the highest-quality version in the editorial sequence and, within versions, the most recent timestamp.

As an additional safeguard, for any firm–event-date pair with multiple transcripts, I retain only the most up-to-date version (i.e., the latest, highest-quality transcript).

Finally, I link firms across data sources by using the Capital IQ mapping table to connect the Capital IQ transcript identifier \texttt{companyId} to Compustat's \texttt{gvkey}.

\clearpage
\section{Prompt Descriptions for Language Generation}
\label{section:prompts}

This appendix documents the set of textual prompts used to systematically modify the tone and content of earnings call transcripts in the experimental framework. Each prompt targets a distinct rhetorical or informational dimension commonly observed in corporate disclosures. The prompts were operationalized using a large language model to simulate variation in managerial communication styles.

\begin{description}

\item[Confidence:] Instructs the model to adopt the tone of a confident and assertive CEO. The prompt removes hedging language and emphasizes decisiveness, authority, and control over business operations and strategy.

\vspace{2mm}
\textit{Prompt:} \\
\texttt{Rewrite the text in the tone of a confident and assertive CEO during an earnings call. Use decisive language, remove hedging expressions, and clearly convey control over business strategy and performance.}

\item[Global Focus:] Reframes the narrative to emphasize macroeconomic conditions, global market dynamics, and high-level industry developments. The prompt links firm-specific performance to broader economic indicators such as inflation, GDP, and regulatory trends.

\vspace{2mm}
\textit{Prompt:} \\
\texttt{Reframe the discussion to emphasize broader macroeconomic trends, global market dynamics, and high-level industry shifts. Draw links between the company's results and wider economic forces such as inflation, GDP, regulatory environments, or sector-wide developments.}

\item[Guidance:] Strengthens forward-looking guidance by prompting the use of clear, strategic, and directional language. The generated text highlights managerial expectations, priorities, and actions while avoiding ambiguity or generic phrasing.

\vspace{2mm}
\textit{Prompt:} \\
\texttt{Clarify and strengthen the company's forward-looking guidance. Use assertive, strategic language to highlight expectations, priorities, and upcoming actions. Where appropriate, specify directionality, expected outcomes, and the rationale behind future decisions. Avoid vague or generic phrasing.}

\item[Sentiment:] Enhances the optimistic tone of the text. The prompt increases the emphasis on business momentum, growth opportunities, and positive outlook, often by highlighting recent wins and expressing confidence in future performance.

\vspace{2mm}
\textit{Prompt:} \\
\texttt{Make the emotional tone more positive and optimistic. Emphasize momentum, exciting opportunities, and enthusiasm about the company's future. Highlight wins and express confidence in continued success.}

\item[Jargon:] Increases the use of technical, domain-specific terminology common in institutional financial communication. Examples include references to ``margin expansion,'' ``revenue mix,'' and ``TAM.'' The prompt maintains semantic fidelity while enriching the financial register.

\vspace{2mm}
\textit{Prompt:} \\
\texttt{Use more technical, domain-specific language typical of institutional earnings calls and financial reporting. Incorporate sector-relevant terminology (e.g., margin expansion, revenue mix, sequential growth, TAM, end-market demand) without changing the underlying meaning or facts.}

\item[Uncertainty:] Prompts the model to introduce or amplify references to business risks and external uncertainties. These may include regulatory pressures, market volatility, or operational dependencies. The tone remains professional but more cautious.

\vspace{2mm}
\textit{Prompt:} \\
\texttt{Highlight and emphasize relevant risks, uncertainties, and operational or market challenges. Introduce cautious language where appropriate, noting external dependencies, regulatory pressures, or volatility in end markets. Maintain a professional tone while acknowledging areas of concern.}

\end{description}

These prompts enable structured variation in narrative tone and content, allowing for controlled analysis of linguistic effects on perception and interpretation of earnings disclosures.

\clearpage
\section{Prompt Descriptions for LLM-as-a-Judge}
\label{section:judge-prompts}

This appendix outlines the prompt template used to evaluate whether a morphed version of an earnings call transcript preserves factual content while exhibiting a significant change in tone.

Each prompt consists of the original (Text A) and the morphed (Text B) version. The large language model is instructed to assess two criteria: (i) that all numerical information and structural elements (e.g., order, scope) are retained, and (ii) that Text B exhibits a clear tonal shift (e.g., more confident or more cautious).

The standardized prompt format is as follows:

\begin{verbatim}
You are an expert in financial communication. 
Below are two excerpts from earnings calls.

Text A: [original excerpt]
Text B: [morphed excerpt]

Does Text B preserve the same numbers and structure as Text A, 
while clearly changing the tone?

Respond with one of the following:
1. Yes: the morphing was executed correctly 
and the language modification is clear and evident,
2. Not sure
3. No: the morphing is inadequate
\end{verbatim}

All evaluations are performed in a zero-shot setting. All responses labeled ``No: the morphing is inadequate'' are discarded from the analysis and reproduced in another instance.

\end{document}